\documentclass[manuscript,screen,review]{acmart}

\AtBeginDocument{%
  \providecommand\BibTeX{{%
    \normalfont B\kern-0.5em{\scshape i\kern-0.25em b}\kern-0.8em\TeX}}}

\setcopyright{acmcopyright}
\copyrightyear{2019}
\acmYear{2019}
\acmDOI{*****}

\usepackage[utf8]{inputenc}
\usepackage[T1]{fontenc}
\usepackage{multirow}

\title{Generating Tertiary Protein Structures via an Interpretative Variational Autoencoder}

\author{Xiaojie Guo}
\affiliation{\institution{Department of Information Sciences and Technology, George Mason University, Fairfax, VA, USA}}

\author{Yuanqi Du}
\affiliation{\institution{Department of Computer Science, George Mason University, Fairfax, VA, USA}}

\author{Sivani Tadepalli}
\affiliation{\institution{Department of Computer Science,George Mason University, Fairfax, VA, USA}}

\author{Liang Zhao}
\affiliation{\institution{Department of Information Sciences and Technology, George Mason University, Fairfax, VA, USA}}
\email{lzhao9@gmu.edu}

\author{Amarda Shehu}
\affiliation{\institution{Department of Computer Science, George Mason University, Fairfax, VA, USA}}
\email{amarda@gmu.edu}


\begin{abstract}
Much scientific enquiry across disciplines is founded upon a mechanistic treatment of dynamic systems that ties form to function. A highly visible instance of this is in molecular biology, where an important goal is to determine functionally-relevant forms/structures that a protein molecule employs to interact with molecular partners in the living cell. This goal is typically pursued under the umbrella of stochastic optimization with algorithms that optimize a scoring function. Research repeatedly shows that current scoring function, though steadily improving, correlate weakly with molecular activity. Inspired by recent momentum in generative deep learning, this paper proposes and evaluates an alternative approach to generating functionally-relevant three-dimensional structures of a protein. Though typically deep generative models struggle with highly-structured data, the work presented here circumvents this challenge via graph-generative models. A comprehensive evaluation of several deep architectures shows the promise of generative models in directly revealing the latent space for sampling novel tertiary structures, as well as in highlighting axes/factors that carry structural meaning and open the black box often associated with deep models. The work presented here is a first step towards interpretative, deep generative models becoming viable and informative complementary approaches to protein structure prediction.

\end{abstract}

\begin{document}
\maketitle
\thispagestyle{empty}
\pagestyle{empty}

\section{Introduction}
\label{sec:Introduction}
Decades of scientific enquiry across disciplines have demonstrated just how fundamental form is to function~\cite{BoehrWright08}. A central question that arises in various scientific domains is how to effectively explore the space of all possible forms of a dynamic system to uncover those that satisfy the constraints needed for the system to be active/functional. In computational structural biology we find a most visible instantiation of this question in the problem of {\it de-novo} (or template-free) protein structure prediction (PSP). PSP seeks to determine one or more biologically-active/native forms/structures of a protein molecule from knowledge of its chemical composition (the sequence of amino acids bonded to one another to form a protein chain)~\cite{LeeZhang17}. 

PSP has a natural and elegant formulation as an optimization problem~\cite{LiwoScheraga99} and is routinely tackled with stochastic optimization algorithms that are tasked with exploring a vast and high-dimensional structure space. These algorithms sample one structure at a time, biased by an expert-designed scoring/energy function that scores sampled structures based on
the likelihood of these strutures being biologically-active/native~\cite{ShehuBookChapter13}. While great progress has been made in stochastic optimization for PSP, there are  fundamental challenges on how to achieve a resource-aware balance between exploration (exploring more of the structure space so as not to miss novel structures) and exploitation (improving structures so as to reach local optima of the scoring function) in vast and high-dimensional search spaces and dealing with inherently-inaccurate scoring functions that often drive towards score-optimal but non-native structures~\cite{KryshtafovichTramontano17}. 

Stochastic optimization frameworks that conduct a biased (by the scoring function) exploration of a protein's structure space, do not learn not to generate structures that are unfit according to the scoring function. Efforts to inject a learning component by Shehu and others struggle with how to best connect the structure generation mechanism/engine in these frameworks with the evaluation engine~\cite{MaximovaShehuPCB16}. A machine learning (ML) framework would seemingly provide a  natural setting; however, while discriminative (ML) models can in principle be built to evaluate tertiary structures, and a growing body of research is showing their promise~\cite{di2012deep, cheng2007improved, eickholt2012predicting, li2011predicting}, effective generative ML models for tertiary protein structures have so far been elusive. 

The majority of work on leveraging deep neural networks (NNs) for problems related to PSP has been on the prediction of distance or contact maps. The latter are alternative representations of tertiary molecular structures that dispense with storing the Cartesian coordinates of atoms in a molecule and instead record only pairwise distances (or, as in contact maps, 0/1 bits to indicate a distance above or not above a threshold). Work on contact map (similarly, distance matrix) prediction focuses on the following setting: given the amino-acid sequence of a protein molecule of interest, predict one contact map that represents the most likely native structure of the target protein. The predicted contact map (or distance matrix) can be utilized to recover a compatible tertiary structure via optimization algorithms that treat the predicted contacts/distances as restraints~\cite{VendruscoloDomany97, adhikari2015confold}. 

Early work on prediction of a contact or distance matrix from a given amino-acid sequence~\cite{KukicPollastri14} employed bidirectional recursive NNs (BRNNs). RaptorX-Contact started a thread of productive research on residual convolutional NNs (CNNs)~\cite{WangXu17}. DNNCON2 leveraged a two-stage CNN~\cite{AdhikariCheng18}. DeepContact deepened the framework via a 9-layer residual CNN with $32$ filters~\cite{LiuPeng18}. DeepCOV alleviated some of the demands for evolutionary information~\cite{JonesKandathil18}. PconsC4 further limited input features and significantly shortens prediction time~\cite{MichelElofsson19}. SPOT-Contact extended RaptorX-Contact by adding a 2D-RNN stage downstream of the CNN~\cite{HansonZhou18}. TripletRes put together four CNNs trained end-to-end~\cite{LiZhang19}.

It is worth emphasizing that the above NNs for contact or pairwise distance prediction predict one instance from a given amino-acid sequence. These models do not generate multiple contact maps or multiple distance matrices to give any view of the diversity of physically-realistic (and possibly functionally-relevant) structures in the structure space of a given protein. The objective of these models is to leverage the strong bias in the input features (which include amino-acid sequence, physico-chemical information of amino acids, co-evolutionary information, secondary structures, solvent accessibility, and other more sophisticated features) to get the model to output \emph{one} prediction that is most likely to represent the native structure. While the performance of these models varies~\cite{TorrisiLe20} and is beyond the scope of this paper, this body of work does not leverage generative deep learning and is not focused on generating a distribution of tertiary structures for a given amino-acid sequence (whether directly in Cartesian space or indirectly through the intermediate representation of contact maps or pairwise distance matrices).

Work in our laboratory has integrated some of these contact prediction models, such as RaptorX-Contact, in optimization-based structure generation engines to evaluate the quality of a sampled tertiary structure in lieu of or in addition to a scoring/energy function\cite{ZamanShehuBCB19}. AlphaFold can also be seen as a similar effort, where a deep NN is used to predict a pairwise distance matrix for a given amino-acid sequence and then predicted distances are encapsulated in a penalty-based scoring function to guide a gradient descent-based optimization algorithm assembling fragments into tertiary structures~\cite{SeniorHassabis19}.

While it remains more challenging to generate a distribution of tertiary structures (whether directly or indirectly through the concept of contact maps or distance matrices), some preliminary efforts have been made~\cite{AnandPossu18,anand2019fully,sabban2019ramanet,ingraham2019learning}. Work in~\cite{AnandPossu18} investigates the promise of such models for protein de-novo design. Other work that focuses on PSP remains preliminary and lack of rigorous evaluation on metrics of interest in PSP and, more generally, protein modeling does not expose what these models have actually learned. -- need to provide more detail here to differentiate these efforts.

Inspired by recent momentum in generative deep learning, we propose here a novel approach based on generative, adversarial deep learning. Leveraging recent opportunities in graph generative learning, we put forth a disentanglement-enhanced contact variational autoencoder (VAE) to which we refer as DECO-VAE from now on. As any other deep neural network, DECO-VAE learns from data, which we obtain from an initial population generated by the state-of-the-art stochastic optimization engine Rosetta~\cite{Leaver-Fay11}. As a VAE, DECO-VAE learns from this data the latent space encapsulating the distribution and then directly generates more data (tertiary structures) from the latent space. 

DECO-VAE addresses several hurdles that stand in the way of generative models for highly-structured data. First, each tertiary structure is represented as a graph via the concept of the contact graph, and DECO-VAE learns the latent space of such graphs, directly sampling novel contact graphs from this space. In this first version, readily-available 3D reconstruction webservers, such as CONFOLD~\cite{adhikari2015confold}, are used to obtain tertiary structures corresponding to DECO-VAE-generated graphs. Second, DECO-VAE addresses the issue of interpretability via the disentanglement mechanism. In one of several insightful analyses presented in this paper, we demonstrate how DECO-VAE is more powerful than a version with no disentanglement mechanism, CO-VAE, as well as other deep architectures. We show that DECO-VAE generates a distribution that does not simply reproduce the input distribution but instead contains novel and better-quality structures. A thorough comparison with several alternative generative models is also presented. In addition, the disentanglement mechanism integrated in DECO-VAE elucidates what exactly about the tertiary structures the "axes"/factors of the learned latent space control, thus opening the black box that is often associated with deep neural networks.

The paper proceeds as follows. First, we present a summary of related work. DECO-VAE, the input data, and other components of our approach (reconstruction of tertiary structures and various analyses and metrics) are then described in detail. The evaluation of the obtained distribution and the interpretability of DECO-VAE is presented next. The paper concludes with a summary of remaining challenges and future opportunities. 

\section{Background and Related Work}
\label{sec:RelatedWork}

The concept of a contact graph or a contact map has long been leveraged in molecular biology, particularly in protein modeling~\cite{wu2008comprehensive}. Briefly, the contact map is an alternatively, compressed representation of a tertiary molecular structure. We can think of it as an adjacenncy matrix, where the entry
$[i, j]$ encodes whether atoms at positions $i, j$ are spatially proximate; the latter uses threshholds, which typically vary in $[8, 10]$\AA in computational biology literature. Typically, the main carbon atoms of each amino acid (the CA atoms) are considered. Some works focus instead on the carbon beta atoms (connecting the side chain atoms to the backbone atoms of an amino acid). In this paper, we take the convention of CA-based contact maps. 

Interest in predicting contacts from a given amino-acid sequence has a long history in protein structure prediction, as it was thought to be perhaps less challenging that predicting other representations (such as Cartesian coordinates). Many machine learning approaches have been used, including
support vector machines~\cite{cheng2007improved,wu2008comprehensive}, random forest~\cite{li2011predicting},
neural networks~\cite{tegge2009nncon,pollastri2002prediction,walsh2009ab,fariselli2001prediction,hamilton2004protein}, and deep learning~\cite{di2012deep, eickholt2012predicting}, with varied success. These approaches struggle with how best to capture the inherent structure in the input training data and how to discover the pertinent latent space.

More recent methods leverage the concept of the latent space via variational autoencoders~\cite{ma2019deep,livi2016generative,anand2018generative}. Specifically, Ma et al.\cite{ma2019deep} develops a deep learning based approach that uses convolutions and a variational autoencoder (a CVAE) to cluster tertiary structures obtained from folding simulations and discover intermediate states from the folding pathways. Anand et al.~\cite{anand2018generative} applies generative adversarial networks (GANs) for {\it de-novo} protein design. The authors encode protein structures in terms of pairwise distances between CA atoms, eliminating the need for the generative
model to learn translational and rotational symmetries. Livi et al.~\cite{livi2016generative} is the first to treat the contact map as a graph and propose a generative model by learning the probability distribution of the existence of a link/contact given the distance between two nodes. 

To the best of our knowledge, no deep graph-generative models have been applied to contact map generation. The DECO-VAE we present in this paper is the first to do so. The method leverages developments in graph-generative neural networks (GNNs). Most of the existing GNN-based graph generation for general graphs have been proposed in the last two years and are based on VAEs~\cite{simonovsky2018graphvae,samanta2019nevae} or GANs~\cite{bojchevski2018netgan}. Most of these approaches generate nodes and edges sequentially to form a whole graph, leading to issues of sensitivity to generation order and high time costs for large graphs. Specifically, GraphVAE~\cite{simonovsky2018graphvae} represents each graph by its adjacent matrix and feature vector and utilizes the VAE model to learn the distribution of the graphs conditioned on a latent representation at the graph level. Graphite~\cite{grover2018graphite} and VGAE~\cite{kipf2016variational} encode the nodes of each graph into node-level embeddings and predict the links between each pair of nodes to generate a graph. In contrast, GraphRNN~\cite{you2018graphrnn} builds an autoregressive generative model on these sequences with a long short-term memory (LSTM) model and shows good scalability. 

These current graph generation models have two drawbacks: (i) the encoder and decoder architecture are not powerful to handle real-world graphs with complex structure relationships; (ii) they are black boxes, lacking interpretability. In the biology domain, we need to understand the features we learn from the observing data and how we generate the contact maps.

In this paper, we additionally address the issue of interpretability via the concept of disentanglement. Recently, a variety of disentangled representation learning algorithms have been proposed for VAEs. Assuming that the data is generated from independent factors of variation (e.g., object orientation and lighting conditions in images of objects), disentanglement as a meta-prior encourages these factors to be captured by different independent variables in the representation. Learning via disentanglement should result in a concise abstract representation of the data useful for a variety of downstream tasks and promises improved sample efficiency. This has motivated a search for ways of learning disentangled representations, where perturbations of an individual dimension of the latent code $z$ perturb the corresponding $x$ in an interpretable manner. A well-known example is the $\beta-VAE$\cite{higgins2017beta}. This has prompted a number of approaches that modify the VAE objective by adding, removing, or altering the weight of individual terms~\cite{kim2018disentangling,chen2018isolating,zhao2017infovae,kumar2017variational,lopez2018information,esmaeili2018structured,alemi2016deep}.

Marrying recent developments in disparate communities, we present here a VAE that generates contact graphs (of tertiary protein structures) and learns disentangled representations that allow interpreting how the learned latent factors control changes at the contact graph level. By utilizing known 3D reconstruction tools, we additionally show how the factors control changes in Cartesian space. The following Section provides further details into the DECO-VAE and other models we design and assess for their ability to generate physically-realistic tertiary protein structures.

\section{Result}
\label{sec:Results}
\subsection*{Comparison Methods}
\label{section: comparison methods}
We utilize three deep graph generation models to validate the superiority of our proposed model. VGAE~\cite{kipf2016variational} is a framework for unsupervised learning on graph-structured data based on the
variational auto-encoder (VAE). This model makes use of latent variables and is capable of learning interpretable latent representations for undirected graphs.  Graphite~\cite{grover2018graphite} is an algorithmic
framework for unsupervised learning of representations over nodes in large graphs using deep latent variable generative models. This model parameterizes variational autoencoders (VAE) with graph neural networks, and uses a novel iterative
graph refinement strategy inspired by low-rank approximations for decoding. GraphRNN~\cite{you2018graphrnn} is a deep autoregressive model to approximate any distribution of graphs with minimal assumptions about their structure. GraphRNN learns to generate graphs by training on a representative set of graphs and decomposes the graph generation process into a sequence of node and edge formations, conditioned on the graph structure generated so far.

\subsection*{Training and Testing Dataset}

The evaluation is carried out on $13$ proteins of varying lengths
($53$ to $146$ amino acids long) and folds ($\alpha$, $\beta$,
$\alpha+\beta$, and $coil$) that are used as a benchmark to evaluate
structure prediction methods~\cite{ZhangZhou18, ZamanShehuBMC19}. On
each protein, given its amino-acid sequence in fasta format, we run the Rosetta AbInitio protocol available in the Rosetta suite~\cite{Leaver-Fay11} to obtain an ensemble of $50,000$ $-$ $60,000$ structures. For each structure, we only retain the central carbon (CA) atom for each amino acid (effectively discarding the other atoms). Each such CA-only structure is converted into a contact graph, as described above. The contact graph dataset of each protein is split into a training and testing data set in a x:y split. The deep models described above are separately trained on each protein.
An overview of Disentanglement-enhanced Contact-VAE (DECO-VAE) is illustrated in Fig.\ref{figure:framework}. By obeserving many decoys contact maps for a protein sequence, it learns the distribution of these decoys and generate many more decoys for that protein sequence. DECO-VAE has two main properties: contact generation and interpretation.
The contact generation is an implementation of a deep graph varitional auto-encoder. The graph encoder learns the distribution of the decoys of a certain sequence and generate new decoys based on a low dimension latent code. And then the contact decoys can be recovered into 3D structure by the existing methods. The contact interpretation ensures each element in the latent representation has a control on an interpretable factor of the generated protein structure. By changing the element value continuously in the latent code, we can control the different properties of the generated protein structure. The proposed method also has a variant which does not consider the disentanglement but only target on the generation of contact maps, named as Contac-VAE (CO-VAE). The CO-VAE has the same structure with the DECO-VAE, except the objective function when training a generative model. The details of the methods are illustrated in the Section of Methods

\subsection*{Comparing the generated contact maps with the native one}
The proposed methods have been evaluated on 13 datasets, each of which corresponds to a protein. For each  protein that is a sequence if amino acids, we use thousands of its contact map decoys to train the CO-VAE and DCO-VAE model. Then based on this well-trained model, the same number of contact maps are generated and compared with the only native one. For each generated contact map and native one, we compute the precision, recall, coverage and and F1 score between them, as shown in Table \ref{table:compare_native}. In addition, the performance of our methods are also compared with the other graph generation methods in machine learning domain (Graphite~\cite{grover2018graphite}, GVAE~\cite{kipf2016variational}, GraphRNN~\cite{you2018graphrnn}). The algorithm and parameter settings of the proposed CO-VAE, DCO-VAE model and the comparison methods are described in details in the Section of Methods.

As shown in Table \ref{table:compare_native}, the graphs generated by CO-VAE and DCO-VAE both have very high F1\_score, and the coverage and recall are all around 0.34, which outperforms the other methods by a large margin. Specifically, our precision is 0.67 and 0.71 on average,  over 53.4\% and 56.7\% higher than the highest performance of comparison methods; our coverage is 0.55 and 0.58 on average,  over 12.7\% and 24.8\% higher than the highest performance of comparison methods; our recall is  0.57 and  0.59 on average,  over 12.2\% and 15.2\% higher than the highest performance of comparison methods; our F1\_score is 0.61 and 0.64 on average,  over 52.4\% and 54.7\% higher than the highest performance of comparison methods; Moreover, the proposed methods (CO-VAE and DCO-VAE) obtain the best performance in four metrics in almost 76\% of the proteins. This indicates that the P-GraphVAE and PD-GraphVAE truly learn the underlined distribution for the contact map decoys for the amino acid sequence. The good performance of the proposed methods rely on its special architecture, where each amino acid has its unique parameters in extracting features from its contacts and the structure similarity is considered among different amino acids in generation process.
\begin{table}[htb]
\scriptsize
\renewcommand\arraystretch{0.9}
  \centering
  \caption{Evaluation of Generated Contact Maps by Comparing to Native One for different proteins}
  \begin{tabular}{llllll|llllll}
   \\\toprule\hline
    Protein&Model &Precision &Coverage &Recall &F1\_score&Protein&Model &Precision &Coverage &Recall &F1\_score \\\hline
        \multirow{5}{*}{1DTJA}& Graphite&0.1320&0.5000&0.5000&0.2031&
        \multirow{5}{*}{1DTDB}& 
      Graphite&0.1526&	0.4893& 0.5002&	0.2338\\
     ~&GVAE&0.1390&0.5109&0.5000&0.2101&~&GVAE&0.1526&	0.4892&	0.5001&	0.2339 \\
    ~&GraphRNN &0.3153&0.1301&0.1301&0.1733&~&GraphRNN &0.2974&	0.2696&	0.2755&	0.2858\\
    ~&CO-VAE &\textbf{0.7432}&0.3310&0.3312&0.4607&~&CO-VAE  &\textbf{0.6144}&	0.5961&	0.6093&	0.6118\\
    ~&DCO-VAE &0.7429&\textbf{0.3347}&\textbf{0.3346}&\textbf{0.4618}&
    ~&DCO-VAE  &0.6136&	\textbf{0.6014}&	\textbf{0.6148}&	\textbf{0.6142}
    \\\hline
     \multirow{5}{*}{1AIL}& 
      Graphite&0.1277&	0.4912&	0.4994&	0.2034& \multirow{5}{*}{2H5ND}& 
      Graphite&0.0748&	0.4976&0.5004&	0.1302\\
     ~&GVAE&0.1276&0.4911&	0.4992&0.2032& ~&GVAE&0.0747&	0.4971&	0.4998&	0.1300 \\
    ~&GraphRNN &0.3900&	0.3718&	0.37802&0.3836&~&GraphRNN &0.2846&	0.2829&0.2845&	0.2844\\
    ~&CO-VAE  &\textbf{0.8244}&	0.8391&0.8531&\textbf{0.8385}&~&CO-VAE  &0.7127&	\textbf{0.6393}&	\textbf{0.6428}&	0.6759\\
    ~&DCO-VAE  &0.8231&	\textbf{0.8393}&\textbf{0.8533}&0.8379&~&DCO-VAE  &\textbf{0.7248}&	0.6354&	0.6389&	\textbf{0.6791}\\\hline
     \multirow{5}{*}{1ALY}& 
      Graphite&0.0684&	0.4981&	0.5002&	0.1203& \multirow{5}{*}{1HHP}& 
      Graphite&0.0902&0.4979&	0.5003&	0.1529\\
     ~&GVAE&0.0683&\textbf{0.4989}&	\textbf{0.5011}&	0.1206&~&GVAE&0.0903&	0.4978&0.5002&	0.1529\\
    ~&GraphRNN &0.2356&	0.1979&	0.1987&	0.2155&~&GraphRNN &0.2804&	0.2582&	0.2594&	0.2693\\
    ~&CO-VAE  &0.8034&	0.4463&	0.4482&	0.5754& ~&CO-VAE &0.6494&	\textbf{0.6612}&\textbf{0.6643}&	0.6567\\
    ~&DCO-VAE &\textbf{0.8368}&	0.4408&	0.4427&	\textbf{0.5791}&~&DCO-VAE  &\textbf{0.6544}&	0.6591&0.6622&	\textbf{0.6583}
\\\hline  
         \multirow{5}{*}{1AOY}& 
      Graphite&0.1065&	0.4963&	0.4995&	0.1756&\multirow{5}{*}{1BQ9}& 
      Graphite&0.1566&0.4809&	0.4994&0.2384\\
     ~&GVAE&0.1066&0.4964&	0.4996&0.1757&~&GVAE&0.1568&	\textbf{0.4815}&	\textbf{0.5001}&	0.2387\\
    ~&GraphRNN &0.3717&	0.2959&	0.2978&	0.3299&~&GraphRNN &0.3659&	0.3232&	0.3356&0.3494\\
    ~&CO-VAE  &0.7620&	\textbf{0.7363}&\textbf{0.7409}&	0.7511&~&CO-VAE  &\textbf{0.4658}&	0.4451&	0.4622&	\textbf{0.4639}\\
    ~&DCO-VAE  &\textbf{0.8109}&	0.7299&	0.7345&	\textbf{0.7708}&~&DCO-VAE  &0.4638&	0.4363&	0.4531&	0.4584
\\\hline  
         \multirow{5}{*}{1C8CA}& 
      Graphite&0.1322&	0.4907&0.5002&	0.2091& \multirow{5}{*}{1CC5}& 
      Graphite&0.1234&0.4843&	0.4997&	0.1980\\
     ~&GVAE&0.1319&	0.4900&0.4995&0.2088&~&GVAE&0.1234&	0.4846&0.4999&	0.1979\\
    ~&GraphRNN &0.3381&	0.3065&	0.3124&	0.3239& ~&GraphRNN &0.3489&	0.2733&	0.2820&	0.3117\\
    ~&CO-VAE  &0.6513&	\textbf{0.6707}&	\textbf{0.6837}&	0.6670&~&CO-VAE  &\textbf{0.8972}&	\textbf{0.5878}&\textbf{0.6064}&	\textbf{0.7237}\\
    ~&DCO-VAE &\textbf{0.6538}&	0.6695&	0.6824&	\textbf{0.6678}&~&DCO-VAE  &0.8965&	0.5864&	0.6051&	0.7225
\\\hline 
         \multirow{5}{*}{1HZ6A}& 
      Graphite&0.1500&\textbf{0.4933}&	\textbf{0.5000}&	0.2308&\multirow{5}{*}{1ISUA}& 
      Graphite&0.1534&	0.4891&0.4996&	0.2348\\
     ~&GVAE&0.1500&	\textbf{0.4933}&	\textbf{0.5000}&	0.2307&~&GVAE&0.1537&	\textbf{0.4898}&	\textbf{0.5003}&	0.2351\\
    ~&GraphRNN &0.1849&	0.3543&	0.3592&	0.2441& ~&GraphRNN &0.2837&	0.2546&	0.2601&	0.2711\\
    ~&CO-VAE &0.5960&	0.4519&	0.4581&	0.5179&~&CO-VAE &0.7542&	0.4738&	0.4839&	0.5886\\
    ~&DCO-VAE &\textbf{0.6067}&	0.4545&0.4607&	\textbf{0.5237}&~&DCO-VAE  &\textbf{0.8435}&	0.4738&	0.4839&	\textbf{0.6151}
\\\hline  
         \multirow{5}{*}{1SAP}& 
      Graphite&0.1329&0.4958&	0.4993&	0.2099& \multirow{5}{*}{1TIG}& 
      Graphite&0.1055&	0.4945&0.5008&	0.1742\\
     ~&GVAE&0.1332&0.4964&0.5000&0.2103& ~&GVAE&0.1054&	0.4939&	0.5001&	0.1740\\
    ~&GraphRNN &0.3021&	0.3178&	0.3202&	0.3107&~&GraphRNN &0.2787&	0.2687&	0.2721&	0.2752\\
    ~&CO-VAE &0.3021&	0.3179&	0.3201&	0.3106&~&CO-VAE &0.5797&	\textbf{0.5691}&	\textbf{0.5764}&	\textbf{0.5780}\\
    ~&DCO-VAE  &\textbf{0.6681}&	\textbf{0.6838}&	\textbf{0.6888}&\textbf{0.6783}&~&DCO-VAE  &\textbf{0.5799}&0.5689&0.5761&	\textbf{0.5780}
\\\hline\hline
     \end{tabular}
  \label{table:compare_native}
\end{table} 
\subsection*{Percentage of native and non-native contacts in a structure}
For each of the generated contact map, we calculate the
percentage of its contacts that are indeed native contacts, namely those contacts in a native structure. Specifically: Given contacts for a structure (each model) and the known native structure, we calculate the percentage of contacts that are in the native structure that are also in the reconstructed structure.  Percentage for each reconstructed structure (model) is calculated for each protein per each method and then an average over all reconstructed structures is reported per protein per method. In addition, we also calculate the percentage of non-native contacts: the contacts that are in a reconstructed structure but not in the structure. For each model, we calculate the number of  non-native contacts and divide it by the number of amino acids in the protein. Then we report the average percentage over all structures per protein per method. The comparison methods are the same ones that mentioned above.

As shown in Table \ref{table:contact evaluation}, the proposed CO-VAE and DCO-VAE both have the highest percentage of the native contacts and the lowest non-native contacts among around 86\% of all the proteins, especially with a very large superiority (e.g. 33\% on average for native contacts and 61\% on average for non-native contacts comparing to the highest performance of comparison methods). The results demonstrate that the proposed methods can accurately generate the contacts like the native one and the generated contact maps have the similar balance between the contacts and non-contacts in the structure. Moreover, even the DCO-VAE enhanced the disentanglement in the objective function of the model, its performance on contact map generation still remains infected, and sometimes even better than the CO-VAE.

\begin{table}[htb]
\small
\renewcommand\arraystretch{0.9}
  \centering
  \caption{Evaluation of the contacts and non-contacts percentage in the generated structures}
  \begin{tabular}{llllllll}\\\hline\hline
    Protein ID&dimension&metric-type&VGAE&Graphite&GraphRNN&	CO-VAE&DCO-VAE\\\hline
    \multirow{2}{*}{1AIL}&73&contacts&49.41&	49.47&	35.94&	\textbf{97.06}&97.00\\
    ~&73&non-contacts&87.86&87.84&62.26&5.14&	\textbf{5.00}\\\hline
    \multirow{2}{*}{1ALY}&146&contacts&49.46&	49.60&	28.86&	\textbf{65.00}&	63.00\\
    ~&146&non-contacts&98.98&98.98&	95.28&	85.07	&\textbf{85.02}\\\hline
        \multirow{2}{*}{1AOY}&78&contacts&49.46&49.44&	27.39&	82.73&	\textbf{83.00}\\
    ~&78&non-contacts&89.39&	89.39&65.10&14.54&\textbf{8.01}\\\hline
        \multirow{2}{*}{1BQ9}&54&contacts&49.57&49.56&	28.64&	\textbf{83.41}&83.34\\
    ~&54&non-contacts&84.57&	84.53&	66.92&	12.73&\textbf{11.43}
\\\hline
        \multirow{2}{*}{1C8CA}&64&contacts&49.44&	49.42&	27.94&92.03&\textbf{93.00}\\
    ~&64&non-contacts&86.68&	86.68&68.11&9.65&\textbf{8.08}\\\hline
        \multirow{2}{*}{1CC5}&83&contacts&49.51&	49.44&	26.49&\textbf{59.83}	&59.52\\
    ~&83&non-contacts&88.66&	88.67&69.09&\textbf{18.62}&	19.06
\\\hline
        \multirow{2}{*}{1DTDB}&61&contacts&49.52&	\textbf{49.57}&	25.65&37.80&38.00\\
    ~&61&non-contacts&86.69&	86.69&74.63&\textbf{65.61}&	66.00
\\\hline
        \multirow{2}{*}{1DTJA}&76&contacts&\textbf{49.56}&	49.49&	10.56&	27.97&	27.02\\
    ~&76&non-contacts&88.18&	88.19&	75.10&	46.86&\textbf{43.71}
\\\hline
        \multirow{2}{*}{1HHP}&99&contacts&49.48&49.55&24.32&90.95&	\textbf{91.11}\\
    ~&99&non-contacts&90.82&90.83&72.88&10.77&\textbf{9.18}\\\hline
        \multirow{2}{*}{1HZ6A}&72&contacts&49.48&	\textbf{49.51}&	35.11&	45.53&	46.00\\
    ~&72&non-contacts&88.50&	88.50&85.53&55.72&	\textbf{55.00}\\\hline
        \multirow{2}{*}{1ISUA}&62&contacts&49.59&	49.43&	24.64&	51.83&	\textbf{52.00}\\
    ~&62&non-contacts&85.15&85.19&72.65&19.65&\textbf{9.34}\\\hline
        \multirow{2}{*}{1SAP}&66&contacts&49.52&	49.43&	29.21&91.70&	\textbf{92.18}\\
    ~&66&non-contacts&86.93&86.96&71.81&\textbf{9.85}&10.47\\\hline
            \multirow{2}{*}{1TIG}&94&contacts&49.75&	50.11&	43.70&	\textbf{100.00}&	\textbf{100.00}\\
    ~&94&non-contacts&99.00&99.00&98.73&\textbf{98.00}&\textbf{98.00}\\\hline
            \multirow{2}{*}{2H5ND}&133&contacts&49.53&49.50&25.22&	82.61&\textbf{81.15}\\
    ~&133&non-contacts&93.56&93.61&	76.86&	19.92&\textbf{19.40}\\\hline\hline
  \end{tabular}\vspace{-0.3cm}
  \label{table:contact evaluation}
\end{table}

\subsection*{Evaluating the learned graph distribution}
Since the successful generation of contact maps rely on successfully learning the distribution of the contact maps, we evaluate whether the generated contact maps follow the learned distributions. By regarding each contact map as a graph, we can calculate four properties of each graphs: density, number of edges, the degree, and transitivity. The density of a graph is the ratio of the number of edges and the number of possible edges. The average degree coefficient measures the similarity of connections in the graph with respect to the node degree. Transitivity is the overall probability for the graph to have adjacent nodes interconnected. All these properties can be calculated by the open source API NetworkX. Then the distance between the distribution of the generated contact maps and the distribution of the training sets in terms of the four properties is measured by three metrics: Pearson correlation coefficient, Bhattacharyya distance and Earth Mover's Distance (EMD). In statistics, the Pearson correlation coefficient (PCC) is a measure of the linear correlation between two variables $X$ and $Y$. Here the $X$ and $Y$ refers to the a kind of graph property of the generated graphs and training graphs respectively. The Bhattacharyya distance and EMD both measures the similarity of two probability distributions and the smaller the better.The results for one example protein are shown as Table \ref{table:distribution evaluation}. In addition, we also compare our proposed methods with the comparison methods. The results for other proteins can be seen in the supplemental materials.

As shown in Table \ref{table:distribution evaluation}, considering EMD distance, the proposed P-GraphVAE has the best performance regarding all the graph properties. For the Bhattacharyya distance, the proposed P-GraphVAE outperformed other comparison methods by 33.38\% and the disentangle version outperformed others by 33.68\%. The Pearson similarity is very small, which demonstrate that the generated graphs are various in some degree than the training set, which ensure the diveristy of the generated contact maps. Specifically, in terms of EMD distance, the proposed CO-VAE and DCO-VAE achieve 1.4 and 2.9 on average, which is 65.8\% and 29.2\% smaller than the best performance comparison method.
\begin{table}[htb]
\small
\renewcommand\arraystretch{0.9}
  \centering
  \caption{Evaluation the learned distributions by comparing to the training sets for Protein 1DTJA\vspace{-0.3cm}}
  \begin{tabular}{lllll}\\\hline\hline
    Graph Property &Methods &Pearson &Bhattacharyya &EMD\\\hline
    \multirow{5}{*}{Density}&Graphite&0.008&5.410&0.465\\
    ~&GVAE &0.007&5.415&0.466\\
    ~&GraphRNN &0.022&\textbf{2.450}&0.004\\
    ~&CO-VAE&0.018&4.941&\textbf{0.002}\\
    ~&DCO-VAE &0.004&3.740&0.004\\\hline
    \multirow{5}{*}{Number of Edge}&Graphite&0.007&5.410&1327\\
    ~& GVAE &0.007&5.415&1329\\
    ~&GraphRNN &0.022&\textbf{2.450}&12.18\\
    ~&CO-VAE&0.019&4.941&\textbf{5.423}  \\
    ~&DCO-VAE &0.004&3.741&11.29\\\hline
     \multirow{5}{*}{Ave-Degree Cor}&Graphite&0.030&5.056&0.136\\
     ~& GVAE &0.009&5.401&0.030\\
     ~&GraphRNN &Nan&3.413&Nan\\
    ~&CO-VAE&0.005&5.310&\textbf{0.017}  \\
    ~&DCO-VAE &0.004&\textbf{3.361}&0.092\\\hline
     \multirow{5}{*}{Transitivity}&Graphite&0.008&5.436&0.244\\
     ~& GVAE &0.008&5.348&0.307\\
     ~&GraphRNN &0.009&\textbf{2.886}&0.122\\
    ~&CO-VAE&0.005&5.159&\textbf{0.004} \\
    ~&DCO-VAE&0.005&3.571&0.027\\\hline\hline
  \end{tabular}
  \label{table:distribution evaluation}
\end{table}

\subsection*{Visualizing the 3D Structure from Generated Contact Maps}
\begin{figure}[htb]
  \centering
    \includegraphics[width=0.95\textwidth]{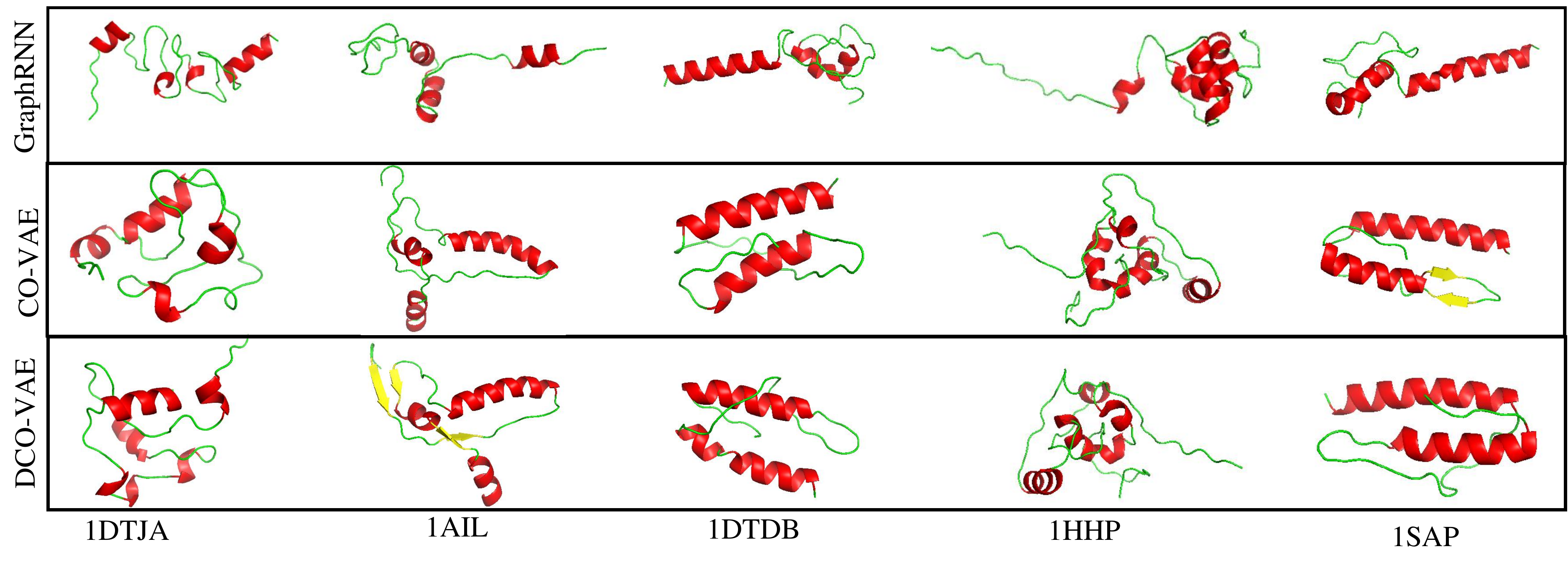}\vspace{-0.5cm}
\caption{Reconstructing the 3D structure from the generated contact maps: we reconstruct the 3D structure based on the generated contact maps of five proteins (i.e.,1DTJA, 1AIL, 1DTDB, 1HHP, and 1SAP) shown as examples; The generated contact maps are generated by the proposed CO-VAE and DCO-VAE, as well as the comparison method GraphRNN. \vspace{-0.3cm}}
\vspace{-0.1cm} 
\label{figure:generated_3D}
\end{figure}
Given the generated contact maps by the proposed CO-VAE and DCO-VAE, we use the existing tool CONFOLD~\cite{adhikari2015confold} to reconstruct the contact map to its responding 3D structure. To visualize and evaluate the real 3D structure of the generated proteins, we also visualize the protein structure by the tool PyMol~\cite{delano2002pymol} which is widely used and user-sponsored molecular visualization system. We show the results both for the proposed CO-VAE model and the DCO-VAE, which is disentangled. And we also compare our proposed model with the graph generative model graphRNN. As shown in Fig. \ref{figure:generated_3D}, the proposed CO-VAE and DCO-VAE have better performance in finding both the secondary and tertiary structure. 

\subsection*{Interpreting the disentangled latent representation}

\begin{figure}[htb]
  \centering
    \includegraphics[width=0.8\textwidth]{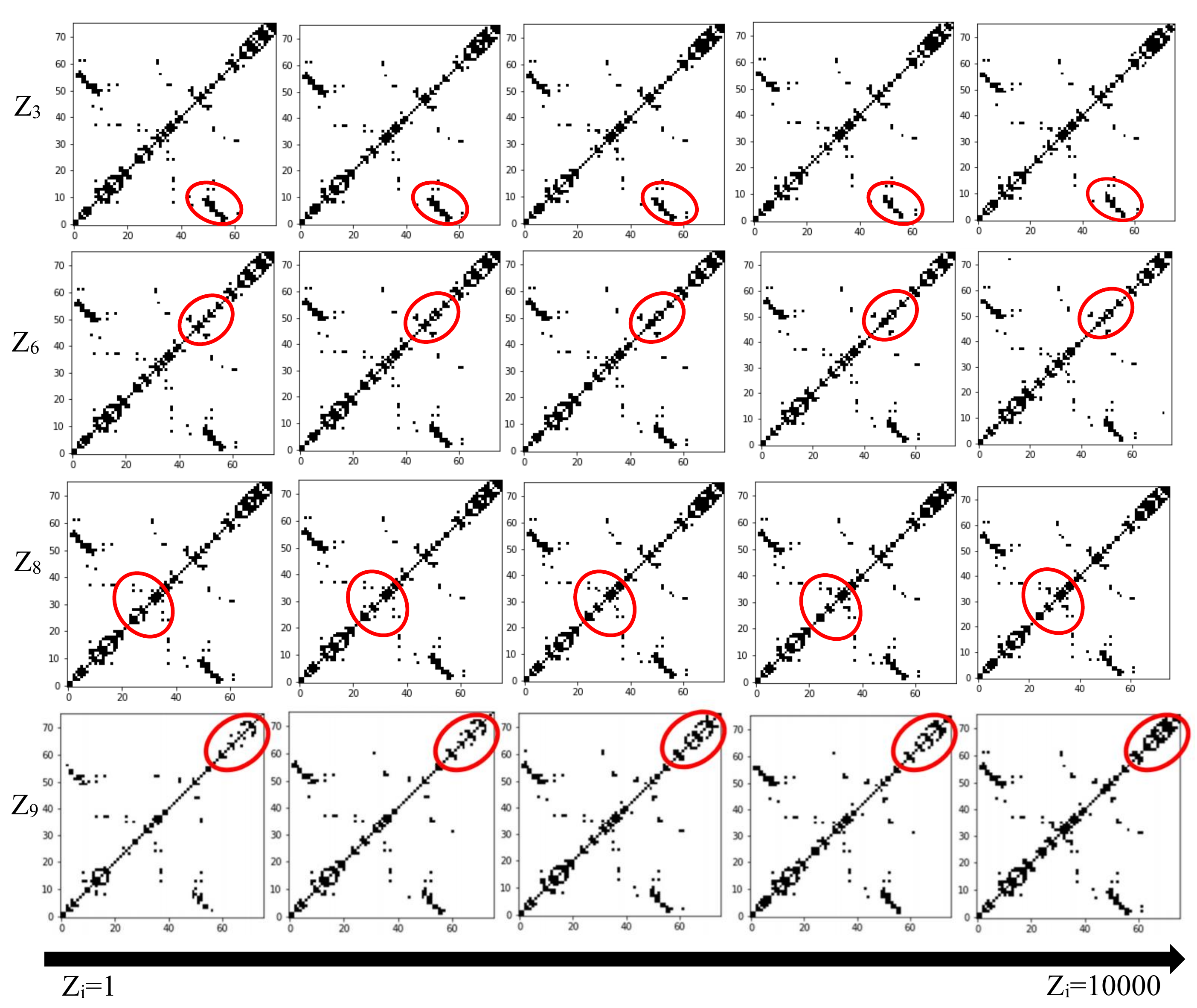}\vspace{-0.3cm}
\caption{Contact map interpretation for protein 1DTJA: Four semantic factors are discovered in the latent variables (i.e., $Z_3$, $Z_6$, $Z_8$, and $Z_9$) which can control the local structural features of the contact maps; the value of latent variables travels from 1 to 10000 and five segments are selected to visualize the local structural feature variations.\vspace{-0.3cm}}
\vspace{-0.1cm} 
\label{figure:interprete_contact map}
\end{figure}

\begin{figure}[htb]
  \centering
    \includegraphics[width=0.8\textwidth]{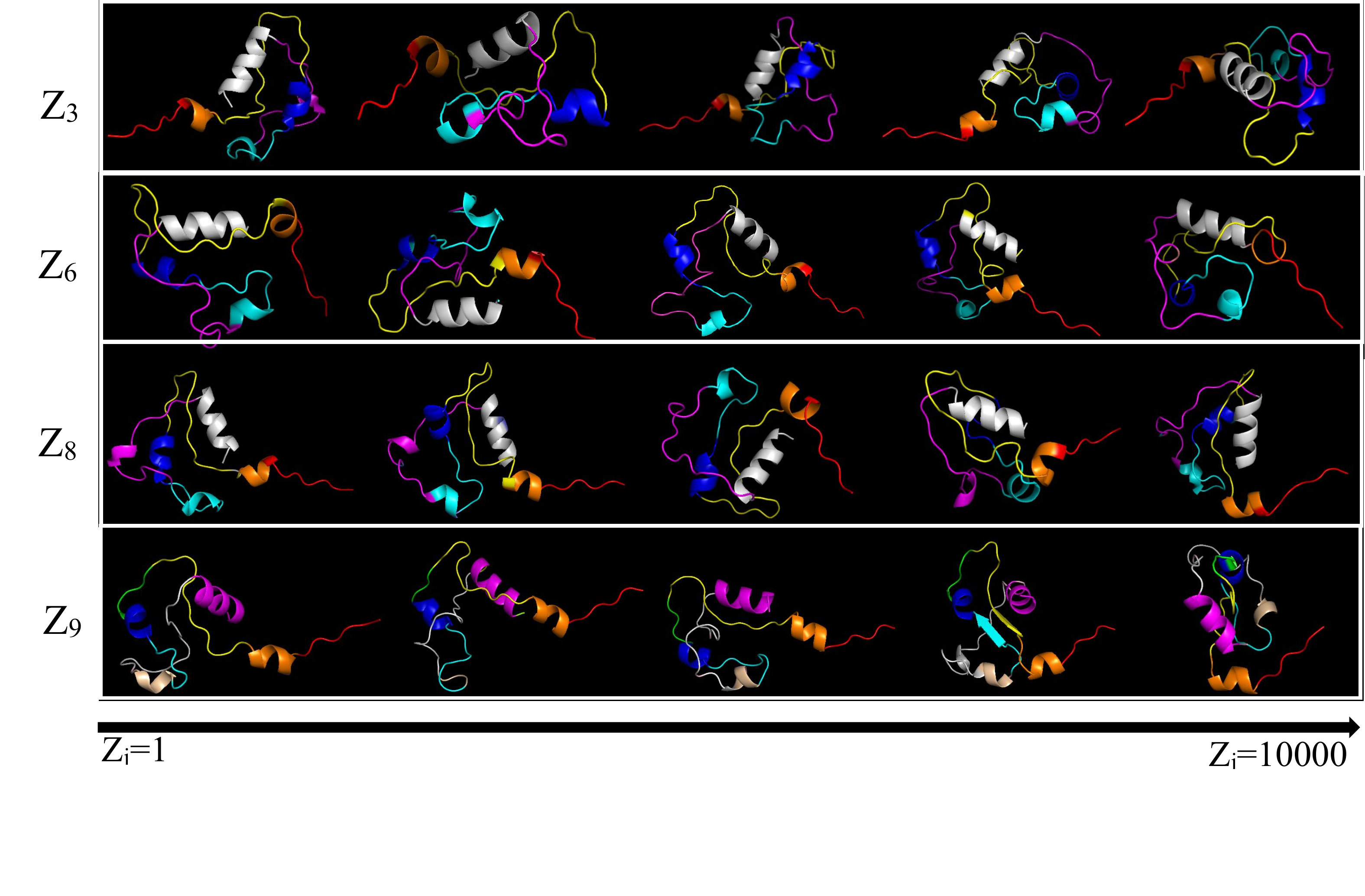}\vspace{-0.5cm}
\caption{Tertiary protein structure interpretation for protein 1DTJA: Tertiary protein structures are reconstructed based on the generated contact maps in Fig.~\ref{figure:interprete_contact map}; the amino acids within the same secondary structure are shown in the same color\vspace{-0.3cm}}
\vspace{-0.1cm} 
\label{figure:interprete}
\end{figure}
It is important to be able to measure the level of disentanglement achieved by different models. In this subsection, we qualitatively demonstrate that our proposed DCO-VAE framework consistently discovers more latent factors and disentangles them in a cleaner fashion. By learning a latent code representation of a protein structure, we assume each variable in the latent code corresponds to a certain factor or property in generating the protein structure. Thus, by changing the value of one variable continuously and fixing the remaining variables, we could visualize the corresponding change of the generated contact map and 3D protein structure. 

First, an example protein IDTJA is analyzed and displayed. the As shown in Fig \ref{figure:interprete} and Fig \ref{figure:interprete_contact map}, we use the convention that in each one latent code varies from left to right while the other latent codes and noise are fixed. The different rows correspond to different random samples of fixed latent codes and noise. For instance, in Fig \ref{figure:interprete_contact map}, one column contains four generated contact map samples regarding each variables, and a row shows the generated contact maps for 5 possible values among 1 and 10000 in a certain variable of latent code with other noise fixed. From the changing contact maps in each row, it is easily to find out which part of the contact map is related or controlled by the certain row. We highlight the part that controlled by each variable in red circle in Fig \ref{figure:interprete_contact map}. The corresponding 3D structure visualization is shown in Fig \ref{figure:interprete_contact map}. On this way, we could easily interpret the role of each variable in the latent code in controlling the generation of the contact maps and the 3D structure. 

\begin{figure}[htb]
  \centering
    \includegraphics[width=0.7\textwidth]{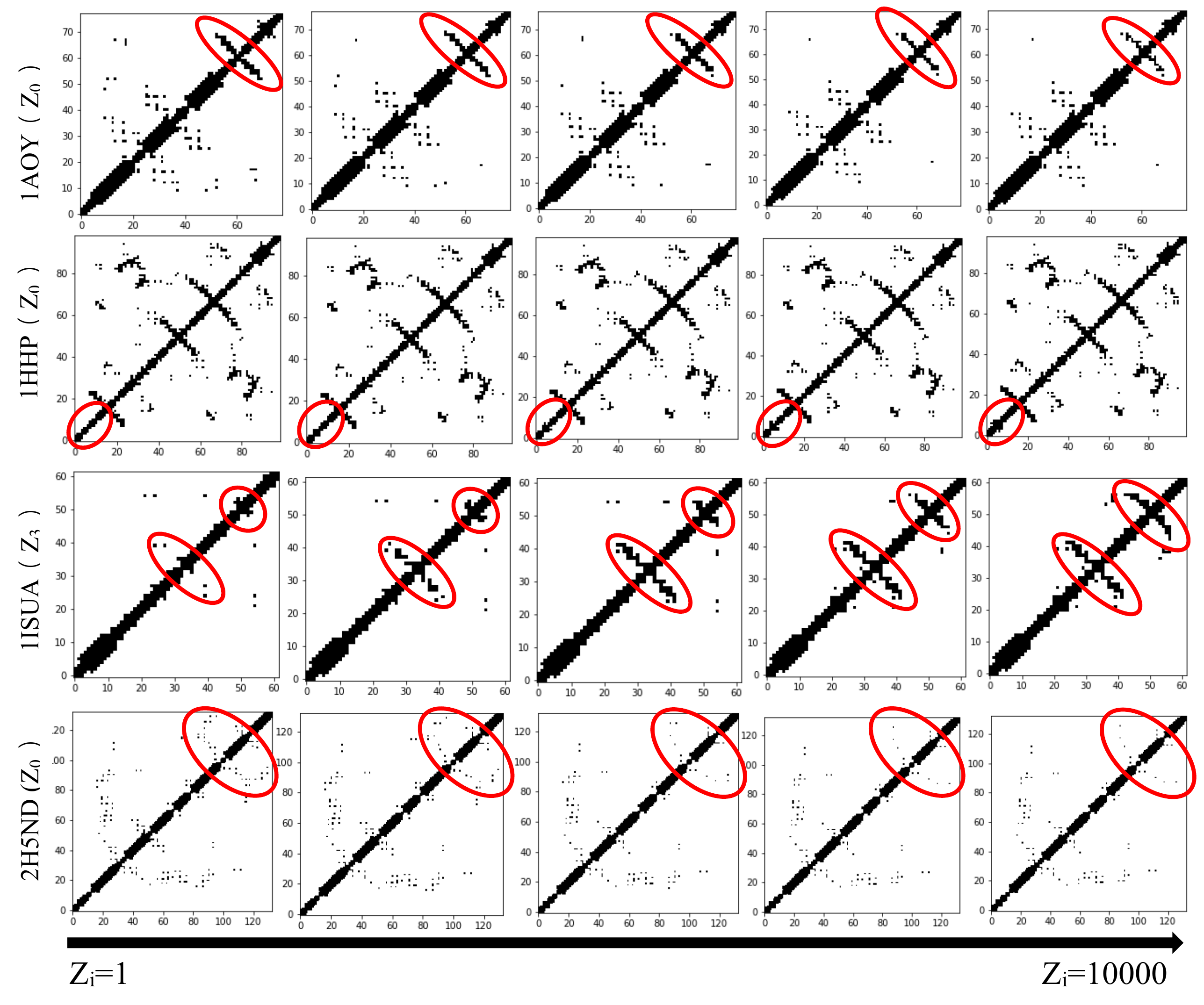}\vspace{-0.3cm}
\caption{Contact map interpretation for different proteins: the generated contact maps of four proteins (i.e.,1AOY, 1HHP, 1ISUA, and 2H5ND) are visualized as examples; for each protein, one of the semantic factors that are discovered in the latent variables is shown to control the local structural features of the contact maps; the value of latent variables travels from 1 to 10000 and five segments are selected to visualize the local structural feature variations.\vspace{-0.3cm}}
\vspace{-0.1cm} 
\label{figure:interprete_contact map2}
\end{figure}

\begin{figure}[htb]
  \centering
    \includegraphics[width=0.8\textwidth]{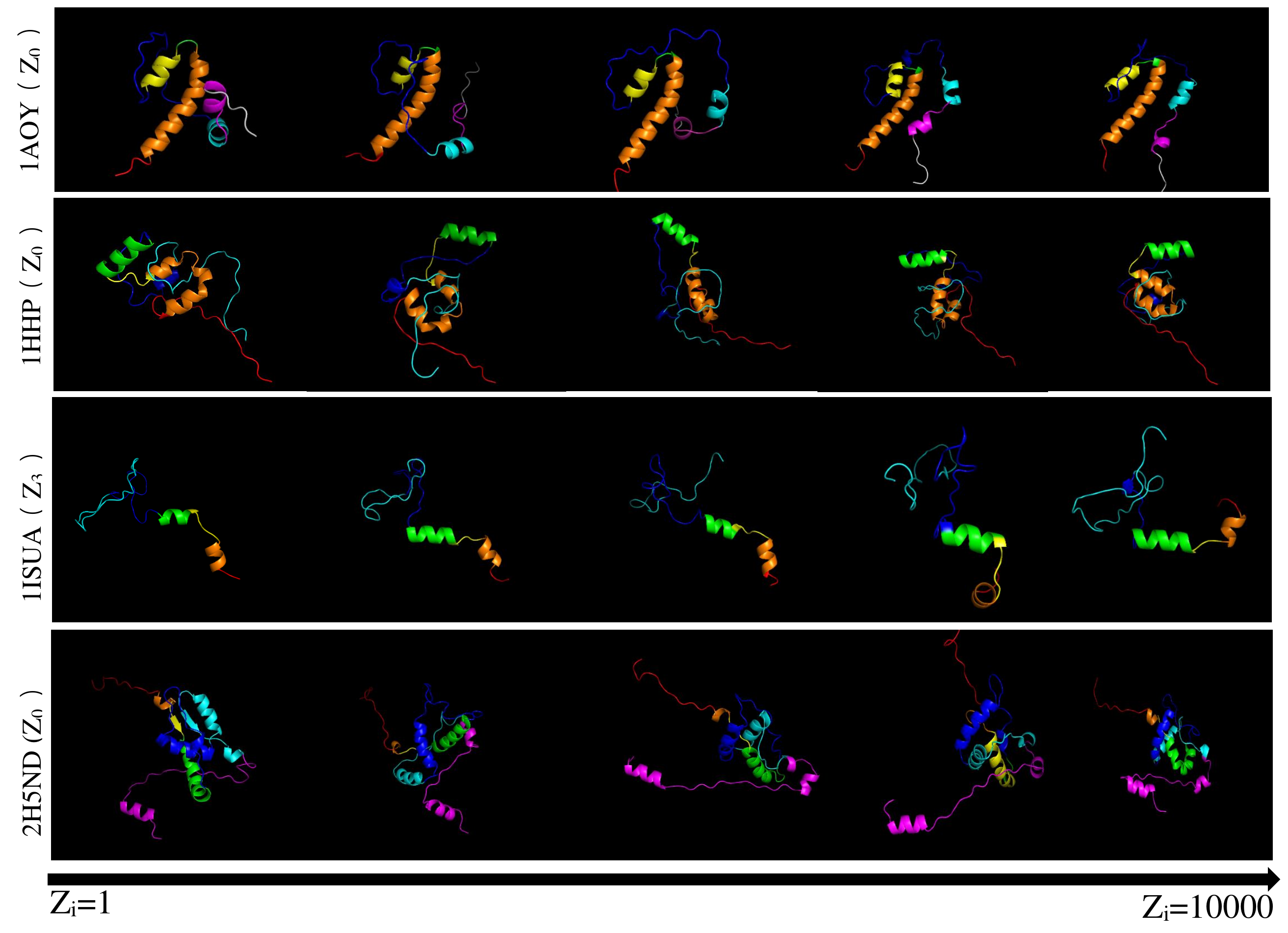}\vspace{-0.3cm}
\caption{Tertiary protein structure interpretation for different proteins: Tertiary protein structures are reconstructed based on the generated contact maps in Fig.~\ref{figure:interprete_contact map2}; the amino acids within the same secondary structure are shown in the same color\vspace{-0.3cm}}
\vspace{-0.1cm} 
\label{figure:interprete2}
\end{figure}

Second, some other proteins are also shown in Fig \ref{figure:interprete2} and Fig \ref{figure:interprete_contact map2}, which demonstrate that the interpretable model can be generalized to any proteins.

\section{Discussion}
\label{sec:Discussion}
In summary, our study aims at utilizing the deep neural network-based generative model for generating tertiary protein structure as well as interpreting the generation process. To the best of our knowledge, we showed for the first time the development and application of an interpretative graph variational auto-encoder for the problem of protein structure generation and interpretation. This demonstrated the promise of generative models in directly revealing the latent space for sampling novel tertiary structures, highlighting axes/factors that carry structural meaning, and opening the black box of deep models.

By treating tertiary protein structure construction problem as the generation of contact maps, we proposed a new disentangled VAE, named Contact Map VAE (CO-VAE) with new graph encoder and decoder. It first learns the underlying distribution of contact maps and then generates additional contact maps by sampling from the learned distribution. The similarity of the structure (e.g., in terms of precision, recall, F1-score, and coverage) as well as positive contact percentage between the generated contact maps and the native ones demonstrated the quality of the generated contact maps. The proposed methods(CO-VAE and DCO-VAE) show great advantages than the existing models by obtaining the best performance in four metrics in almost 76\% of the proteins and the highest percentage of the native contacts as well as the lowest non-native contacts among around 86\% of all the proteins. Furthermore, we reconstructed the generated contact maps into 3D protein structures. The visualization of the constructed 3D protein illustrated that the generated contact maps are valid and useful for further generating the 3D tertiary protein structure. 

To further investigate whether the proposed CO-VAE indeed learns the distribution of the observed contact map samples, we generated a set of contact maps and compared their underlying distribution and that of the real contact maps. Since it is difficult to directly evaluate the graph distribution, we resorted to evaluating the distribution of the properties of graphs, such as density, edge numbers, average degree correlation, and transitivity. The small EMD and Bhattachryya distances between the learned and real distribution in terms of all four properties of graphs generated by CO-VAE and DCO-VAE validated that the underlying graph distribution is effectively discovered, which outperformed the distances calculated from the graphs that generated by comparison methods by 33.4\% and 47.5\% on average. Though the Pearson correlation score was very low, it showed that the diversity of the generated graphs was ensured since it measures the strength and direction of a linear relationship between two variables other than the similarity of two distributions. 

Next, to explore our generative model's capability of interpreting the process of contact map generation, we enhanced our CO-VAE by enforcing the weights of the  second term of the training objective, leading to the disentanglement among variables and hence a new interpretable model named DCO-VAE. The learned latent representation variables in DCO-VAE is expected to be related to the factors that influence the formation of the contact maps. As a result, for each latent variable, by varying the value of this variable from 1 to 10,000 and fixing the values of others at the same time, a certain local part of the generated contact maps showed obvious trends (e.g., contracting or sketching). This demonstrated that the learned latent variables are effectively disentangled, which indicated the potential semantic factors in the formation of contact maps and the corresponding protein structures.    

From one aspect, though there have been some deep generative models applied to the protein contact map generation, they cannot utilize the relationships among amino acids by treating the contact maps as the graph-structured data with graph generative models. From another aspect, though many interpretable learning models have been explored and applied into the image generation, there are no interpretable models that can discover the semantic factors that can control the process of protein folding. In summary, to the best of our knowledge, this is the first time an interpretable deep generative model for graphs have been applied for protein structure prediction and its effectiveness in generating good-quality protein structures is demonstrated. In addition, the proposed DCO-VAE can also be applied to other real-world applications where there is a need for graph-structured data generation and interpretation. 

There are much more promising and challenging topics that can be originated from the research in this paper for future exploration. For example, it would be interesting and potentially beneficial to develop an end-to-end tertiary protein structure generation model directly for 3D structure instead of contact maps. This is because there is a gap between the contact map generation and 3D structure formulation process, and the learned variables can only explain well of the formation of contact maps instead of the 3D structure. In addition, the exploration of the node-edge joint generative model would also be highly interesting. The proposed DCO-VAE model focuses on generating the graph topology (i.e., via contact maps) instead of node features (e.g., properties and types of amino acid). Jointly generating both graph topology and node features could be important in some cases, such as when directly generating the 3D structure where the node features can be the 3D positions. 

\section{Methods}
\label{sec:Methods}

\subsection*{Problem Formulation}

\begin{figure}[htb]
  \centering
    \includegraphics[width=0.9\textwidth]{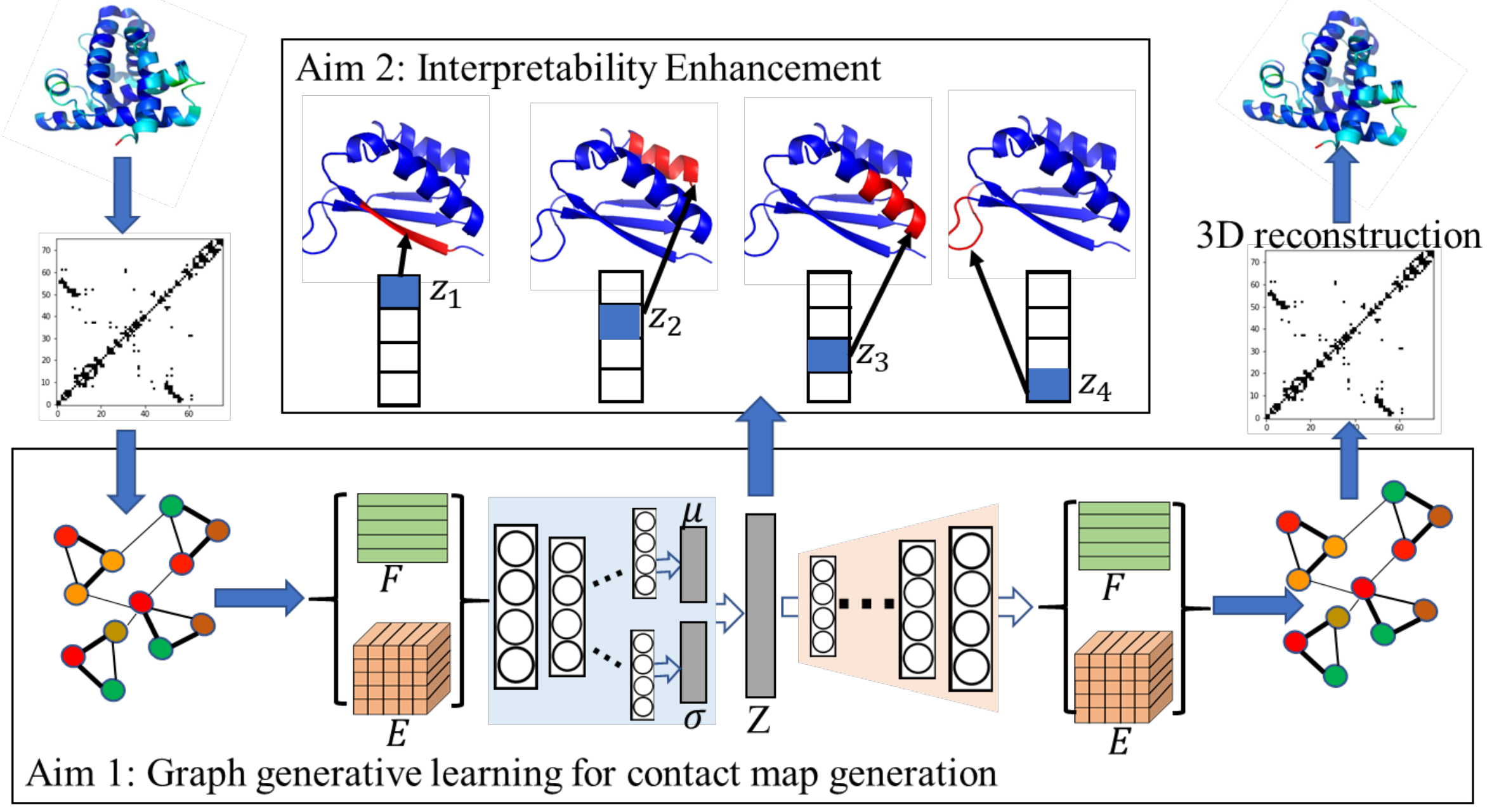}\vspace{-0.3cm}
\caption{Overall schematic of proposed generative learning framework.\vspace{-0.3cm}}
\label{figure:framework}
\end{figure}

Recoverable tertiary structure formulation requires to preserve information not only on which atom is bonded with which else, but, more importantly, on which atoms are in proximity of each-other in three-dimensional space. As shown in Fig.~\ref{figure:framework}, to address this issue, we employ contact map graph which can be trivially computed
from tertiary structures as the input to our model. And to recover a tertiary structure from a contact
map is comparatively trivial~\cite{adhikari2015confold}. 

Hence, the contact map is a graph-based representation of a tertiary structure that effectively embeds the three-dimensional Cartesian space into
a two-dimensional space. Specifically, in our approach, let the contact graph $G = (E,F)$ now be associated with its edge attribute tensor $E\in R^{N\times N \times L_1}$, which  denotes the adjacent matrix $A$ when $L_1=1$; and node attribute matrix $F\in R^{N\times L_2}$ (as shown in Fig.~\ref{figure:framework}) denoting the identity of each atom by hot vector embedding, where $N$ is the number of atoms over which contacts are computed; one can do so for all atoms in a molecule, or representative atoms to control the size of the input space. The edge and node attributes are rich mechanisms to add additional information. For instance, node attributes can store not just the identities of the amino acids but also their PSSM profile (derived from the Position Specific Scoring Matrix), as well as their solvent accessibility and secondary structure as derived from a given tertiary structure. The edge attributes can encode additional information about contacts, such as their exact distance and/or contact predicted for that pair of amino acids (vertices) from sequence information alone. In our experiment, we are given an undirected, unweighted graph $G$ as mentioned above, and we only use the adjacency matrix $A$ of $G$ (we assume diagonal elements set to 1, i.e. every node is connected to itself) and node attributes $F$ store the identities of the amino acids with $L_2=20$ (the number of all the atom types).

\subsection*{Deep Contact Map Variational Auto-encoder}
\label{section:graphVAE}

The task in this paper requires to learn the disentangled generative models on contact map that encodes graph information. Although disentanglement enhancement and deep graph generative models are respectively attracting fast-increasing attention in recent years, the synergistic integration of them has rarely been explored yet. To learn the distribution of the contact map decoys of a amino acid sequence, we first propose a new graph varational auto-encoder framework that consists of powerful graph encoder, graph decoder, and latent graph code disentanglement mechanism. First we introduce the model without disentanglement mechanism, named as The Deep Contact Map Variational Auto-encoder (CO-VAE).

Specifically, given the adjacent matrix $A$ of a contact map, we further introduce stochastic latent variables $z_i$, summarized in an $1\times H$ vector $Z$. The overall architecture contains a graph encoder and decoder which are trained by optimizing the variational lower bound Loss w.r.t. the variational parameters $W_i$:
\begin{equation}
    \mathcal{L}=\mathbf{E}_{q(Z|F,A)}[logp(A|Z,F)]-KL[q(Z|F,A)||p(Z)]
    \label{loss}
\end{equation}
where $KL[q(\cdot)||p(\cdot)]$ is the Kullback-Leibler divergence between $q(\cdot)$ and $p(\cdot)$. The first item is the reconstruction loss of the generated contact maps and the second term enforces the inferenced latent vector close to the real latent vector distribution. We take a Gaussian prior $p(Z) =\prod_i p(z_i)=\prod_i \mathcal{N}(z_i|0,I)$ 

For the encoder part, we take a simple inference model parameterized by a two-layer Graph Convolution Nueral Network (GCN)~\cite{kipf2016semi} as the encoder of our CO-VAE:
\begin{equation}
    q(Z|F,A)=\prod_{i=1}^{N}q(z_i|F,A), \quad where \quad q(z_i|F,A)=\mathcal{N}(z_i|\mu_i,diag(\sigma^2))
\end{equation}
Here $\mu=GCN_\mu(F,A)$ is the mean of the latent vectors, which is inferenced by GCN; similarly $log\sigma=GCN_\sigma(F,A)$ is the standard of the latent vector that is inferenced by anotehr GCN. Thus, it is possible to sample $Z$ from the distributions of the latent vectors $q(Z|F,A)$.   

For the generative model, we utilize the graph decoder network proposed in our previous works\cite{guo2018deep}. This work first proposes the graph deconvolution based graph decoders, which achieve the best performance in the graph generation task when the node set of the graph is fixed. Thus, we choose this graph decoder as part of our CO-VAE.
\begin{equation}
    p(A|Z,F)=\prod_{j=1}^{N}\prod_{i=1}^{N}p(A_{ij}|Z)
\end{equation}
where $A_{ij}$ are the elements of $A$. 

We perform full-batch gradient descent and make use of the re-parameterization trick~\cite{kingma2013auto} for training. The architecture and mathematical operations of the graph encoder for modelling the $q(Z|A,F)$ and decoder for modelling the $p(A|Z,F)$ are detailed in the supplemental materials.

\subsection*{Disentangled Contact Map VAE}
Then next we introduce the Contact Map VAE with the disentanglement mechanism, which is inpired by the $\beta-VAE$~\cite{higgins2017beta}.
$\beta-VAE$ is proposed for
automated discovery of interpretable factorised latent representations from data in a completely unsupervised manner, and has been currently used in image domain~\cite{huang2018multimodal} and Natural Language Processing~\cite{yi2018neural}. It is a modification of the variational autoencoder (VAE) framework by introducing an adjustable hyperparameter $\beta$ that balances latent channel capacity and independence constraints
with reconstruction accuracy. Thus, we proposed the Disentangled Contact Map VAE (DCO-VAE) to largely interpret the latent representations for the protein generation.
We propose augmenting the CO-VAE framework with a single hyperparameter $\beta$ that modulates the learning constraints applied to the model. These constraints impose a limit on the capacity of the latent information channel and control the emphasis on learning statistically independent latent factors. Eq.~\ref{loss} can be re-written to arrive at the DCO-VAE formulation, but with the addition of the $\beta$ coefficient:
\begin{equation}
    \mathcal{L}=\mathbf{E}_{q(Z|F,A)}[logp(A|Z,F)]-\beta KL[q(Z|F,A)||p(Z)]
    \label{disen_loss}
\end{equation}
where with the $\beta=1$ corresponds to the CO-VAE framework mentioned above; and with $\beta>1$ the model is pushed to learn a more efficient latent representation of the data, which is disentangled if the data contains at least some underlying factors of variation that are independent.

\section*{Data availability}
All data, models, and source code are freely available to readers upon contacting the authors.

\bibliographystyle{ACM-Reference-Format}
\bibliography{main}


\begin{thebibliography}{59}


\ifx \showCODEN    \undefined \def \showCODEN     #1{\unskip}     \fi
\ifx \showDOI      \undefined \def \showDOI       #1{#1}\fi
\ifx \showISBNx    \undefined \def \showISBNx     #1{\unskip}     \fi
\ifx \showISBNxiii \undefined \def \showISBNxiii  #1{\unskip}     \fi
\ifx \showISSN     \undefined \def \showISSN      #1{\unskip}     \fi
\ifx \showLCCN     \undefined \def \showLCCN      #1{\unskip}     \fi
\ifx \shownote     \undefined \def \shownote      #1{#1}          \fi
\ifx \showarticletitle \undefined \def \showarticletitle #1{#1}   \fi
\ifx \showURL      \undefined \def \showURL       {\relax}        \fi
\providecommand\bibfield[2]{#2}
\providecommand\bibinfo[2]{#2}
\providecommand\natexlab[1]{#1}
\providecommand\showeprint[2][]{arXiv:#2}

\bibitem[\protect\citeauthoryear{Adhikari, Bhattacharya, Cao, and
  Cheng}{Adhikari et~al\mbox{.}}{2015}]%
        {adhikari2015confold}
\bibfield{author}{\bibinfo{person}{Badri Adhikari}, \bibinfo{person}{Debswapna
  Bhattacharya}, \bibinfo{person}{Renzhi Cao}, {and} \bibinfo{person}{Jianlin
  Cheng}.} \bibinfo{year}{2015}\natexlab{}.
\newblock \showarticletitle{CONFOLD: residue-residue contact-guided ab initio
  protein folding}.
\newblock \bibinfo{journal}{\emph{Proteins: Structure, Function, and
  Bioinformatics}} \bibinfo{volume}{83}, \bibinfo{number}{8}
  (\bibinfo{year}{2015}), \bibinfo{pages}{1436--1449}.
\newblock


\bibitem[\protect\citeauthoryear{Adhikari, Hou, and Cheng}{Adhikari
  et~al\mbox{.}}{2018}]%
        {AdhikariCheng18}
\bibfield{author}{\bibinfo{person}{B. Adhikari}, \bibinfo{person}{J. Hou},
  {and} \bibinfo{person}{J. Cheng}.} \bibinfo{year}{2018}\natexlab{}.
\newblock \showarticletitle{{DNCON2:} improved protein contact prediction using
  two-level deep convolutional neural networks}.
\newblock \bibinfo{journal}{\emph{Bioinformatics}}  \bibinfo{volume}{34}
  (\bibinfo{year}{2018}), \bibinfo{pages}{1466--1472}.
\newblock


\bibitem[\protect\citeauthoryear{Alemi, Fischer, Dillon, and Murphy}{Alemi
  et~al\mbox{.}}{2016}]%
        {alemi2016deep}
\bibfield{author}{\bibinfo{person}{Alexander~A Alemi}, \bibinfo{person}{Ian
  Fischer}, \bibinfo{person}{Joshua~V Dillon}, {and} \bibinfo{person}{Kevin
  Murphy}.} \bibinfo{year}{2016}\natexlab{}.
\newblock \showarticletitle{Deep variational information bottleneck}.
\newblock \bibinfo{journal}{\emph{arXiv preprint arXiv:1612.00410}}
  (\bibinfo{year}{2016}).
\newblock


\bibitem[\protect\citeauthoryear{Anand, Eguchi, and Huang}{Anand
  et~al\mbox{.}}{2019}]%
        {anand2019fully}
\bibfield{author}{\bibinfo{person}{Namrata Anand}, \bibinfo{person}{Raphael
  Eguchi}, {and} \bibinfo{person}{Po-Ssu Huang}.}
  \bibinfo{year}{2019}\natexlab{}.
\newblock \showarticletitle{Fully differentiable full-atom protein backbone
  generation}.
\newblock  (\bibinfo{year}{2019}).
\newblock


\bibitem[\protect\citeauthoryear{Anand and Huang}{Anand and Huang}{2018a}]%
        {AnandPossu18}
\bibfield{author}{\bibinfo{person}{Namrata Anand} {and} \bibinfo{person}{Possu
  Huang}.} \bibinfo{year}{2018}\natexlab{a}.
\newblock \showarticletitle{Generative modeling for protein structures}. In
  \bibinfo{booktitle}{\emph{Advances in Neural Information Processing
  Systems}}. \bibinfo{pages}{7494--7505}.
\newblock


\bibitem[\protect\citeauthoryear{Anand and Huang}{Anand and Huang}{2018b}]%
        {anand2018generative}
\bibfield{author}{\bibinfo{person}{Namrata Anand} {and} \bibinfo{person}{Possu
  Huang}.} \bibinfo{year}{2018}\natexlab{b}.
\newblock \showarticletitle{Generative modeling for protein structures}. In
  \bibinfo{booktitle}{\emph{Advances in Neural Information Processing
  Systems}}. \bibinfo{pages}{7494--7505}.
\newblock


\bibitem[\protect\citeauthoryear{Boehr and Wright}{Boehr and Wright}{2008}]%
        {BoehrWright08}
\bibfield{author}{\bibinfo{person}{D.~D. Boehr} {and} \bibinfo{person}{P.~E.
  Wright}.} \bibinfo{year}{2008}\natexlab{}.
\newblock \showarticletitle{How do proteins interact?}
\newblock \bibinfo{journal}{\emph{Science}} \bibinfo{volume}{320},
  \bibinfo{number}{5882} (\bibinfo{year}{2008}), \bibinfo{pages}{1429--1430}.
\newblock


\bibitem[\protect\citeauthoryear{Bojchevski, Shchur, Z{\"u}gner, and
  G{\"u}nnemann}{Bojchevski et~al\mbox{.}}{2018}]%
        {bojchevski2018netgan}
\bibfield{author}{\bibinfo{person}{Aleksandar Bojchevski},
  \bibinfo{person}{Oleksandr Shchur}, \bibinfo{person}{Daniel Z{\"u}gner},
  {and} \bibinfo{person}{Stephan G{\"u}nnemann}.}
  \bibinfo{year}{2018}\natexlab{}.
\newblock \showarticletitle{Netgan: Generating graphs via random walks}.
\newblock \bibinfo{journal}{\emph{arXiv preprint arXiv:1803.00816}}
  (\bibinfo{year}{2018}).
\newblock


\bibitem[\protect\citeauthoryear{Chen, Li, Grosse, and Duvenaud}{Chen
  et~al\mbox{.}}{2018}]%
        {chen2018isolating}
\bibfield{author}{\bibinfo{person}{Tian~Qi Chen}, \bibinfo{person}{Xuechen Li},
  \bibinfo{person}{Roger~B Grosse}, {and} \bibinfo{person}{David~K Duvenaud}.}
  \bibinfo{year}{2018}\natexlab{}.
\newblock \showarticletitle{Isolating sources of disentanglement in variational
  autoencoders}. In \bibinfo{booktitle}{\emph{Advances in Neural Information
  Processing Systems}}. \bibinfo{pages}{2610--2620}.
\newblock


\bibitem[\protect\citeauthoryear{Cheng and Baldi}{Cheng and Baldi}{2007}]%
        {cheng2007improved}
\bibfield{author}{\bibinfo{person}{Jianlin Cheng} {and} \bibinfo{person}{Pierre
  Baldi}.} \bibinfo{year}{2007}\natexlab{}.
\newblock \showarticletitle{Improved residue contact prediction using support
  vector machines and a large feature set}.
\newblock \bibinfo{journal}{\emph{BMC bioinformatics}} \bibinfo{volume}{8},
  \bibinfo{number}{1} (\bibinfo{year}{2007}), \bibinfo{pages}{113}.
\newblock


\bibitem[\protect\citeauthoryear{DeLano}{DeLano}{2002}]%
        {delano2002pymol}
\bibfield{author}{\bibinfo{person}{Warren~Lyford DeLano}.}
  \bibinfo{year}{2002}\natexlab{}.
\newblock \bibinfo{title}{PyMOL}.
\newblock
\newblock


\bibitem[\protect\citeauthoryear{Di~Lena, Nagata, and Baldi}{Di~Lena
  et~al\mbox{.}}{2012}]%
        {di2012deep}
\bibfield{author}{\bibinfo{person}{Pietro Di~Lena}, \bibinfo{person}{Ken
  Nagata}, {and} \bibinfo{person}{Pierre Baldi}.}
  \bibinfo{year}{2012}\natexlab{}.
\newblock \showarticletitle{Deep architectures for protein contact map
  prediction}.
\newblock \bibinfo{journal}{\emph{Bioinformatics}} \bibinfo{volume}{28},
  \bibinfo{number}{19} (\bibinfo{year}{2012}), \bibinfo{pages}{2449--2457}.
\newblock


\bibitem[\protect\citeauthoryear{Eickholt and Cheng}{Eickholt and
  Cheng}{2012}]%
        {eickholt2012predicting}
\bibfield{author}{\bibinfo{person}{Jesse Eickholt} {and}
  \bibinfo{person}{Jianlin Cheng}.} \bibinfo{year}{2012}\natexlab{}.
\newblock \showarticletitle{Predicting protein residue--residue contacts using
  deep networks and boosting}.
\newblock \bibinfo{journal}{\emph{Bioinformatics}} \bibinfo{volume}{28},
  \bibinfo{number}{23} (\bibinfo{year}{2012}), \bibinfo{pages}{3066--3072}.
\newblock


\bibitem[\protect\citeauthoryear{Esmaeili, Wu, Jain, Bozkurt, Siddharth, Paige,
  Brooks, Dy, and van~de Meent}{Esmaeili et~al\mbox{.}}{2018}]%
        {esmaeili2018structured}
\bibfield{author}{\bibinfo{person}{Babak Esmaeili}, \bibinfo{person}{Hao Wu},
  \bibinfo{person}{Sarthak Jain}, \bibinfo{person}{Alican Bozkurt},
  \bibinfo{person}{Narayanaswamy Siddharth}, \bibinfo{person}{Brooks Paige},
  \bibinfo{person}{Dana~H Brooks}, \bibinfo{person}{Jennifer Dy}, {and}
  \bibinfo{person}{Jan-Willem van~de Meent}.} \bibinfo{year}{2018}\natexlab{}.
\newblock \showarticletitle{Structured disentangled representations}.
\newblock \bibinfo{journal}{\emph{arXiv preprint arXiv:1804.02086}}
  (\bibinfo{year}{2018}).
\newblock


\bibitem[\protect\citeauthoryear{Fariselli, Olmea, Valencia, and
  Casadio}{Fariselli et~al\mbox{.}}{2001}]%
        {fariselli2001prediction}
\bibfield{author}{\bibinfo{person}{Piero Fariselli}, \bibinfo{person}{Osvaldo
  Olmea}, \bibinfo{person}{Alfonso Valencia}, {and} \bibinfo{person}{Rita
  Casadio}.} \bibinfo{year}{2001}\natexlab{}.
\newblock \showarticletitle{Prediction of contact maps with neural networks and
  correlated mutations}.
\newblock \bibinfo{journal}{\emph{Protein engineering}} \bibinfo{volume}{14},
  \bibinfo{number}{11} (\bibinfo{year}{2001}), \bibinfo{pages}{835--843}.
\newblock


\bibitem[\protect\citeauthoryear{Grover, Zweig, and Ermon}{Grover
  et~al\mbox{.}}{2018}]%
        {grover2018graphite}
\bibfield{author}{\bibinfo{person}{Aditya Grover}, \bibinfo{person}{Aaron
  Zweig}, {and} \bibinfo{person}{Stefano Ermon}.}
  \bibinfo{year}{2018}\natexlab{}.
\newblock \showarticletitle{Graphite: Iterative generative modeling of graphs}.
\newblock \bibinfo{journal}{\emph{arXiv preprint arXiv:1803.10459}}
  (\bibinfo{year}{2018}).
\newblock


\bibitem[\protect\citeauthoryear{Guo, Wu, and Zhao}{Guo et~al\mbox{.}}{2018}]%
        {guo2018deep}
\bibfield{author}{\bibinfo{person}{Xiaojie Guo}, \bibinfo{person}{Lingfei Wu},
  {and} \bibinfo{person}{Liang Zhao}.} \bibinfo{year}{2018}\natexlab{}.
\newblock \showarticletitle{Deep Graph Translation}.
\newblock \bibinfo{journal}{\emph{arXiv preprint arXiv:1805.09980}}
  (\bibinfo{year}{2018}).
\newblock


\bibitem[\protect\citeauthoryear{Hamilton, Burrage, Ragan, and Huber}{Hamilton
  et~al\mbox{.}}{2004}]%
        {hamilton2004protein}
\bibfield{author}{\bibinfo{person}{Nicholas Hamilton}, \bibinfo{person}{Kevin
  Burrage}, \bibinfo{person}{Mark~A Ragan}, {and} \bibinfo{person}{Thomas
  Huber}.} \bibinfo{year}{2004}\natexlab{}.
\newblock \showarticletitle{Protein contact prediction using patterns of
  correlation}.
\newblock \bibinfo{journal}{\emph{Proteins: Structure, Function, and
  Bioinformatics}} \bibinfo{volume}{56}, \bibinfo{number}{4}
  (\bibinfo{year}{2004}), \bibinfo{pages}{679--684}.
\newblock


\bibitem[\protect\citeauthoryear{Hanson, Paliwal, Litfin, Yang, and
  Zhou}{Hanson et~al\mbox{.}}{2018}]%
        {HansonZhou18}
\bibfield{author}{\bibinfo{person}{J. Hanson}, \bibinfo{person}{K. Paliwal},
  \bibinfo{person}{T. Litfin}, \bibinfo{person}{Y. Yang}, {and}
  \bibinfo{person}{Y. Zhou}.} \bibinfo{year}{2018}\natexlab{}.
\newblock \showarticletitle{Accurate prediction of protein contact maps by
  coupling residual two-dimensional bidirectional long short-term memory with
  convolutional neural networks}.
\newblock \bibinfo{journal}{\emph{Bioinformatics}}  \bibinfo{volume}{34}
  (\bibinfo{year}{2018}), \bibinfo{pages}{4039--4045}.
\newblock


\bibitem[\protect\citeauthoryear{Higgins, Matthey, Pal, Burgess, Glorot,
  Botvinick, Mohamed, and Lerchner}{Higgins et~al\mbox{.}}{2017}]%
        {higgins2017beta}
\bibfield{author}{\bibinfo{person}{Irina Higgins}, \bibinfo{person}{Loic
  Matthey}, \bibinfo{person}{Arka Pal}, \bibinfo{person}{Christopher Burgess},
  \bibinfo{person}{Xavier Glorot}, \bibinfo{person}{Matthew Botvinick},
  \bibinfo{person}{Shakir Mohamed}, {and} \bibinfo{person}{Alexander
  Lerchner}.} \bibinfo{year}{2017}\natexlab{}.
\newblock \showarticletitle{beta-VAE: Learning Basic Visual Concepts with a
  Constrained Variational Framework.}
\newblock \bibinfo{journal}{\emph{ICLR}} \bibinfo{volume}{2},
  \bibinfo{number}{5} (\bibinfo{year}{2017}), \bibinfo{pages}{6}.
\newblock


\bibitem[\protect\citeauthoryear{Huang, Liu, Belongie, and Kautz}{Huang
  et~al\mbox{.}}{2018}]%
        {huang2018multimodal}
\bibfield{author}{\bibinfo{person}{Xun Huang}, \bibinfo{person}{Ming-Yu Liu},
  \bibinfo{person}{Serge Belongie}, {and} \bibinfo{person}{Jan Kautz}.}
  \bibinfo{year}{2018}\natexlab{}.
\newblock \showarticletitle{Multimodal unsupervised image-to-image
  translation}. In \bibinfo{booktitle}{\emph{Proceedings of the European
  Conference on Computer Vision (ECCV)}}. \bibinfo{pages}{172--189}.
\newblock


\bibitem[\protect\citeauthoryear{Ingraham, Riesselman, Sander, and
  Marks}{Ingraham et~al\mbox{.}}{2019}]%
        {ingraham2019learning}
\bibfield{author}{\bibinfo{person}{John Ingraham}, \bibinfo{person}{Adam
  Riesselman}, \bibinfo{person}{Chris Sander}, {and} \bibinfo{person}{Debora
  Marks}.} \bibinfo{year}{2019}\natexlab{}.
\newblock \showarticletitle{Learning protein structure with a differentiable
  simulator}. In \bibinfo{booktitle}{\emph{International Conference on Learning
  Representations}}.
\newblock


\bibitem[\protect\citeauthoryear{Jones and Kandathil}{Jones and
  Kandathil}{2018}]%
        {JonesKandathil18}
\bibfield{author}{\bibinfo{person}{D.~T. Jones} {and} \bibinfo{person}{S.~M.
  Kandathil}.} \bibinfo{year}{2018}\natexlab{}.
\newblock \showarticletitle{High precision in protein contact prediction using
  fully convolutional neural networks and minimal sequence features}.
\newblock \bibinfo{journal}{\emph{Bioinformatics}}  \bibinfo{volume}{34}
  (\bibinfo{year}{2018}), \bibinfo{pages}{3308--3315}.
\newblock


\bibitem[\protect\citeauthoryear{Kim and Mnih}{Kim and Mnih}{2018}]%
        {kim2018disentangling}
\bibfield{author}{\bibinfo{person}{Hyunjik Kim} {and} \bibinfo{person}{Andriy
  Mnih}.} \bibinfo{year}{2018}\natexlab{}.
\newblock \showarticletitle{Disentangling by factorising}.
\newblock \bibinfo{journal}{\emph{arXiv preprint arXiv:1802.05983}}
  (\bibinfo{year}{2018}).
\newblock


\bibitem[\protect\citeauthoryear{Kingma and Welling}{Kingma and
  Welling}{2013}]%
        {kingma2013auto}
\bibfield{author}{\bibinfo{person}{Diederik~P Kingma} {and}
  \bibinfo{person}{Max Welling}.} \bibinfo{year}{2013}\natexlab{}.
\newblock \showarticletitle{Auto-encoding variational bayes}.
\newblock \bibinfo{journal}{\emph{arXiv preprint arXiv:1312.6114}}
  (\bibinfo{year}{2013}).
\newblock


\bibitem[\protect\citeauthoryear{Kipf and Welling}{Kipf and Welling}{2016a}]%
        {kipf2016semi}
\bibfield{author}{\bibinfo{person}{Thomas~N Kipf} {and} \bibinfo{person}{Max
  Welling}.} \bibinfo{year}{2016}\natexlab{a}.
\newblock \showarticletitle{Semi-supervised classification with graph
  convolutional networks}.
\newblock \bibinfo{journal}{\emph{arXiv preprint arXiv:1609.02907}}
  (\bibinfo{year}{2016}).
\newblock


\bibitem[\protect\citeauthoryear{Kipf and Welling}{Kipf and Welling}{2016b}]%
        {kipf2016variational}
\bibfield{author}{\bibinfo{person}{Thomas~N Kipf} {and} \bibinfo{person}{Max
  Welling}.} \bibinfo{year}{2016}\natexlab{b}.
\newblock \showarticletitle{Variational graph auto-encoders}.
\newblock \bibinfo{journal}{\emph{arXiv preprint arXiv:1611.07308}}
  (\bibinfo{year}{2016}).
\newblock


\bibitem[\protect\citeauthoryear{Kryshtafovych, Monastyrskyy, Fidelis, Schwede,
  and Tramontano}{Kryshtafovych et~al\mbox{.}}{2017}]%
        {KryshtafovichTramontano17}
\bibfield{author}{\bibinfo{person}{A. Kryshtafovych}, \bibinfo{person}{B.
  Monastyrskyy}, \bibinfo{person}{K. Fidelis}, \bibinfo{person}{T. Schwede},
  {and} \bibinfo{person}{A. Tramontano}.} \bibinfo{year}{2017}\natexlab{}.
\newblock \showarticletitle{Assessment of model accuracy estimations in
  {CASP12}}.
\newblock \bibinfo{journal}{\emph{Proteins: Struct, Funct, Bioinf}}
  \bibinfo{volume}{86}, \bibinfo{number}{Suppl 1} (\bibinfo{year}{2017}),
  \bibinfo{pages}{345--360}.
\newblock


\bibitem[\protect\citeauthoryear{Kukic, Mirabello, Tradigo, Walsh, Veltri, and
  Pollastri}{Kukic et~al\mbox{.}}{2014}]%
        {KukicPollastri14}
\bibfield{author}{\bibinfo{person}{P. Kukic}, \bibinfo{person}{P. Mirabello},
  \bibinfo{person}{G. Tradigo}, \bibinfo{person}{I. Walsh}, \bibinfo{person}{P.
  Veltri}, {and} \bibinfo{person}{G. Pollastri}.}
  \bibinfo{year}{2014}\natexlab{}.
\newblock \showarticletitle{Toward an accurate prediction of inter-residue
  distances in proteins using 2d recursive neural networks}.
\newblock \bibinfo{journal}{\emph{BMC Bioinf}}  \bibinfo{volume}{15}
  (\bibinfo{year}{2014}), \bibinfo{pages}{6}.
\newblock


\bibitem[\protect\citeauthoryear{Kumar, Sattigeri, and Balakrishnan}{Kumar
  et~al\mbox{.}}{2017}]%
        {kumar2017variational}
\bibfield{author}{\bibinfo{person}{Abhishek Kumar}, \bibinfo{person}{Prasanna
  Sattigeri}, {and} \bibinfo{person}{Avinash Balakrishnan}.}
  \bibinfo{year}{2017}\natexlab{}.
\newblock \showarticletitle{Variational inference of disentangled latent
  concepts from unlabeled observations}.
\newblock \bibinfo{journal}{\emph{arXiv preprint arXiv:1711.00848}}
  (\bibinfo{year}{2017}).
\newblock


\bibitem[\protect\citeauthoryear{Leaver-Fay, Tyka, Lewis, Lange, Thompson,
  Jacak, et~al\mbox{.}}{Leaver-Fay et~al\mbox{.}}{2011}]%
        {Leaver-Fay11}
\bibfield{author}{\bibinfo{person}{A. Leaver-Fay}, \bibinfo{person}{M. Tyka},
  \bibinfo{person}{S.~M. Lewis}, \bibinfo{person}{O.~F. Lange},
  \bibinfo{person}{J. Thompson}, \bibinfo{person}{R. Jacak}, {et~al\mbox{.}}}
  \bibinfo{year}{2011}\natexlab{}.
\newblock \showarticletitle{{ROSETTA3:} an object-oriented software suite for
  the simulation and design of macromolecules}.
\newblock \bibinfo{journal}{\emph{Methods Enzymol}}  \bibinfo{volume}{487}
  (\bibinfo{year}{2011}), \bibinfo{pages}{545--574}.
\newblock


\bibitem[\protect\citeauthoryear{Lee, Freddolino, and Zhang}{Lee
  et~al\mbox{.}}{2017}]%
        {LeeZhang17}
\bibfield{author}{\bibinfo{person}{J. Lee}, \bibinfo{person}{P. Freddolino},
  {and} \bibinfo{person}{Y. Zhang}.} \bibinfo{year}{2017}\natexlab{}.
\newblock \showarticletitle{Ab initio protein structure prediction}.
\newblock In \bibinfo{booktitle}{\emph{From Protein Structure to Function with
  Bioinformatics} (\bibinfo{edition}{2} ed.)},
  \bibfield{editor}{\bibinfo{person}{D.~J. Rigden}} (Ed.).
  \bibinfo{publisher}{Springer London}, Chapter~1, \bibinfo{pages}{3--35}.
\newblock


\bibitem[\protect\citeauthoryear{Li, Fang, and Fang}{Li et~al\mbox{.}}{2011}]%
        {li2011predicting}
\bibfield{author}{\bibinfo{person}{Yunqi Li}, \bibinfo{person}{Yaping Fang},
  {and} \bibinfo{person}{Jianwen Fang}.} \bibinfo{year}{2011}\natexlab{}.
\newblock \showarticletitle{Predicting residue--residue contacts using random
  forest models}.
\newblock \bibinfo{journal}{\emph{Bioinformatics}} \bibinfo{volume}{27},
  \bibinfo{number}{24} (\bibinfo{year}{2011}), \bibinfo{pages}{3379--3384}.
\newblock


\bibitem[\protect\citeauthoryear{Li, Zhang, Bell, Yu, and Zhang}{Li
  et~al\mbox{.}}{2019}]%
        {LiZhang19}
\bibfield{author}{\bibinfo{person}{Y. Li}, \bibinfo{person}{C. Zhang},
  \bibinfo{person}{E.~W. Bell}, \bibinfo{person}{{D.-J.} Yu}, {and}
  \bibinfo{person}{Y. Zhang}.} \bibinfo{year}{2019}\natexlab{}.
\newblock \showarticletitle{Ensembling multiple raw coevolutionary features
  with deep residual neural networks for contact-map prediction in {CASP13}}.
\newblock \bibinfo{journal}{\emph{Proteins: Struct, Funct, Bioinf}}
  \bibinfo{volume}{87}, \bibinfo{number}{12} (\bibinfo{year}{2019}),
  \bibinfo{pages}{1082--1091}.
\newblock


\bibitem[\protect\citeauthoryear{Liu, Palmedo, Q., Berger, and Peng}{Liu
  et~al\mbox{.}}{2018}]%
        {LiuPeng18}
\bibfield{author}{\bibinfo{person}{Y. Liu}, \bibinfo{person}{P. Palmedo},
  \bibinfo{person}{Ye Q.}, \bibinfo{person}{B. Berger}, {and}
  \bibinfo{person}{J. Peng}.} \bibinfo{year}{2018}\natexlab{}.
\newblock \showarticletitle{Enhancing evolutionary couplings with deep
  convolutional neural networks}.
\newblock \bibinfo{journal}{\emph{Cell Syst}} \bibinfo{volume}{6},
  \bibinfo{number}{e3} (\bibinfo{year}{2018}), \bibinfo{pages}{65--74}.
\newblock


\bibitem[\protect\citeauthoryear{Livi, Maiorino, Giuliani, Rizzi, and
  Sadeghian}{Livi et~al\mbox{.}}{2016}]%
        {livi2016generative}
\bibfield{author}{\bibinfo{person}{Lorenzo Livi}, \bibinfo{person}{Enrico
  Maiorino}, \bibinfo{person}{Alessandro Giuliani}, \bibinfo{person}{Antonello
  Rizzi}, {and} \bibinfo{person}{Alireza Sadeghian}.}
  \bibinfo{year}{2016}\natexlab{}.
\newblock \showarticletitle{A generative model for protein contact networks}.
\newblock \bibinfo{journal}{\emph{Journal of Biomolecular Structure and
  Dynamics}} \bibinfo{volume}{34}, \bibinfo{number}{7} (\bibinfo{year}{2016}),
  \bibinfo{pages}{1441--1454}.
\newblock


\bibitem[\protect\citeauthoryear{Liwo, Lee, Ripoll, Pillardy, and
  Scheraga}{Liwo et~al\mbox{.}}{1999}]%
        {LiwoScheraga99}
\bibfield{author}{\bibinfo{person}{A. Liwo}, \bibinfo{person}{J. Lee},
  \bibinfo{person}{D.~R. Ripoll}, \bibinfo{person}{J. Pillardy}, {and}
  \bibinfo{person}{H.~A. Scheraga}.} \bibinfo{year}{1999}\natexlab{}.
\newblock \showarticletitle{Protein structure prediction by global optimization
  of a potential energy function}.
\newblock \bibinfo{journal}{\emph{Proc Natl Acad Sci USA}}
  \bibinfo{volume}{96}, \bibinfo{number}{10} (\bibinfo{year}{1999}),
  \bibinfo{pages}{5482--5485}.
\newblock


\bibitem[\protect\citeauthoryear{Lopez, Regier, Jordan, and Yosef}{Lopez
  et~al\mbox{.}}{2018}]%
        {lopez2018information}
\bibfield{author}{\bibinfo{person}{Romain Lopez}, \bibinfo{person}{Jeffrey
  Regier}, \bibinfo{person}{Michael~I Jordan}, {and} \bibinfo{person}{Nir
  Yosef}.} \bibinfo{year}{2018}\natexlab{}.
\newblock \showarticletitle{Information constraints on auto-encoding
  variational bayes}. In \bibinfo{booktitle}{\emph{Advances in Neural
  Information Processing Systems}}. \bibinfo{pages}{6114--6125}.
\newblock


\bibitem[\protect\citeauthoryear{Ma, Bhowmik, Lee, Turilli, Young, Jha, and
  Ramanathan}{Ma et~al\mbox{.}}{2019}]%
        {ma2019deep}
\bibfield{author}{\bibinfo{person}{Heng Ma}, \bibinfo{person}{Debsindhu
  Bhowmik}, \bibinfo{person}{Hyungro Lee}, \bibinfo{person}{Matteo Turilli},
  \bibinfo{person}{Michael~T Young}, \bibinfo{person}{Shantenu Jha}, {and}
  \bibinfo{person}{Arvind Ramanathan}.} \bibinfo{year}{2019}\natexlab{}.
\newblock \showarticletitle{Deep generative model driven protein folding
  simulation}.
\newblock \bibinfo{journal}{\emph{arXiv preprint arXiv:1908.00496}}
  (\bibinfo{year}{2019}).
\newblock


\bibitem[\protect\citeauthoryear{Maximova, Moffatt, Ma, Nussinov, and
  Shehu}{Maximova et~al\mbox{.}}{2016}]%
        {MaximovaShehuPCB16}
\bibfield{author}{\bibinfo{person}{T. Maximova}, \bibinfo{person}{R. Moffatt},
  \bibinfo{person}{B. Ma}, \bibinfo{person}{R. Nussinov}, {and}
  \bibinfo{person}{A. Shehu}.} \bibinfo{year}{2016}\natexlab{}.
\newblock \showarticletitle{Principles and Overview of Sampling Methods for
  Modeling Macromolecular Structure and Dynamics}.
\newblock \bibinfo{journal}{\emph{{PLoS} Comp. Biol.}} \bibinfo{volume}{12},
  \bibinfo{number}{4} (\bibinfo{year}{2016}), \bibinfo{pages}{e1004619}.
\newblock


\bibitem[\protect\citeauthoryear{Michel, Hurtado, and Elofsson}{Michel
  et~al\mbox{.}}{2019}]%
        {MichelElofsson19}
\bibfield{author}{\bibinfo{person}{M. Michel}, \bibinfo{person}{D.~M. Hurtado},
  {and} \bibinfo{person}{A. Elofsson}.} \bibinfo{year}{2019}\natexlab{}.
\newblock \showarticletitle{PconsC4: fast, accurate and hassle-free contact
  predictions}.
\newblock \bibinfo{journal}{\emph{Bioinformatics}}  \bibinfo{volume}{35}
  (\bibinfo{year}{2019}), \bibinfo{pages}{2677--2679}.
\newblock


\bibitem[\protect\citeauthoryear{Pollastri and Baldi}{Pollastri and
  Baldi}{2002}]%
        {pollastri2002prediction}
\bibfield{author}{\bibinfo{person}{Gianluca Pollastri} {and}
  \bibinfo{person}{Pierre Baldi}.} \bibinfo{year}{2002}\natexlab{}.
\newblock \showarticletitle{Prediction of contact maps by GIOHMMs and recurrent
  neural networks using lateral propagation from all four cardinal corners}.
\newblock \bibinfo{journal}{\emph{Bioinformatics}} \bibinfo{volume}{18},
  \bibinfo{number}{suppl\_1} (\bibinfo{year}{2002}), \bibinfo{pages}{S62--S70}.
\newblock


\bibitem[\protect\citeauthoryear{Sabban and Markovsky}{Sabban and
  Markovsky}{2019}]%
        {sabban2019ramanet}
\bibfield{author}{\bibinfo{person}{Sari Sabban} {and} \bibinfo{person}{Mikhail
  Markovsky}.} \bibinfo{year}{2019}\natexlab{}.
\newblock \showarticletitle{RamaNet: Computational De Novo Protein Design using
  a Long Short-Term Memory Generative Adversarial Neural Network}.
\newblock \bibinfo{journal}{\emph{BioRxiv}} (\bibinfo{year}{2019}),
  \bibinfo{pages}{671552}.
\newblock


\bibitem[\protect\citeauthoryear{Samanta, Abir, Jana, Chattaraj, Ganguly, and
  Rodriguez}{Samanta et~al\mbox{.}}{2019}]%
        {samanta2019nevae}
\bibfield{author}{\bibinfo{person}{Bidisha Samanta}, \bibinfo{person}{DE Abir},
  \bibinfo{person}{Gourhari Jana}, \bibinfo{person}{Pratim~Kumar Chattaraj},
  \bibinfo{person}{Niloy Ganguly}, {and} \bibinfo{person}{Manuel~Gomez
  Rodriguez}.} \bibinfo{year}{2019}\natexlab{}.
\newblock \showarticletitle{Nevae: A deep generative model for molecular
  graphs}. In \bibinfo{booktitle}{\emph{Proceedings of the AAAI Conference on
  Artificial Intelligence}}, Vol.~\bibinfo{volume}{33}.
  \bibinfo{pages}{1110--1117}.
\newblock


\bibitem[\protect\citeauthoryear{Senior, Evans, Jumper, Kirkpatrick, Sifre,
  et~al\mbox{.}}{Senior et~al\mbox{.}}{2019}]%
        {SeniorHassabis19}
\bibfield{author}{\bibinfo{person}{A.~W. Senior}, \bibinfo{person}{R. Evans},
  \bibinfo{person}{J. Jumper}, \bibinfo{person}{J. Kirkpatrick},
  \bibinfo{person}{L. Sifre}, {et~al\mbox{.}}} \bibinfo{year}{2019}\natexlab{}.
\newblock \showarticletitle{Protein structure prediction using multiple deep
  neural networks in {CASP13}}.
\newblock \bibinfo{journal}{\emph{Proteins: Struct, Funct, Bioinf}}
  \bibinfo{volume}{87}, \bibinfo{number}{12} (\bibinfo{year}{2019}),
  \bibinfo{pages}{1141--1148}.
\newblock


\bibitem[\protect\citeauthoryear{Shehu}{Shehu}{2013}]%
        {ShehuBookChapter13}
\bibfield{author}{\bibinfo{person}{A. Shehu}.} \bibinfo{year}{2013}\natexlab{}.
\newblock \showarticletitle{Probabilistic Search and Optimization for Protein
  Energy Landscapes}.
\newblock In \bibinfo{booktitle}{\emph{Handbook of Computational Molecular
  Biology}}, \bibfield{editor}{\bibinfo{person}{S.~Aluru} {and}
  \bibinfo{person}{A.~Singh}} (Eds.). \bibinfo{publisher}{Chapman \& Hall/CRC
  Computer \& Information Science Series}.
\newblock


\bibitem[\protect\citeauthoryear{Simonovsky and Komodakis}{Simonovsky and
  Komodakis}{2018}]%
        {simonovsky2018graphvae}
\bibfield{author}{\bibinfo{person}{Martin Simonovsky} {and}
  \bibinfo{person}{Nikos Komodakis}.} \bibinfo{year}{2018}\natexlab{}.
\newblock \showarticletitle{Graphvae: Towards generation of small graphs using
  variational autoencoders}. In \bibinfo{booktitle}{\emph{International
  Conference on Artificial Neural Networks}}. Springer,
  \bibinfo{pages}{412--422}.
\newblock


\bibitem[\protect\citeauthoryear{Tegge, Wang, Eickholt, and Cheng}{Tegge
  et~al\mbox{.}}{2009}]%
        {tegge2009nncon}
\bibfield{author}{\bibinfo{person}{Allison~N Tegge}, \bibinfo{person}{Zheng
  Wang}, \bibinfo{person}{Jesse Eickholt}, {and} \bibinfo{person}{Jianlin
  Cheng}.} \bibinfo{year}{2009}\natexlab{}.
\newblock \showarticletitle{NNcon: improved protein contact map prediction
  using 2D-recursive neural networks}.
\newblock \bibinfo{journal}{\emph{Nucleic acids research}}
  \bibinfo{volume}{37}, \bibinfo{number}{suppl\_2} (\bibinfo{year}{2009}),
  \bibinfo{pages}{W515--W518}.
\newblock


\bibitem[\protect\citeauthoryear{Torrisi, Pollastri, and Le}{Torrisi
  et~al\mbox{.}}{2020}]%
        {TorrisiLe20}
\bibfield{author}{\bibinfo{person}{M. Torrisi}, \bibinfo{person}{G. Pollastri},
  {and} \bibinfo{person}{Q. Le}.} \bibinfo{year}{2020}\natexlab{}.
\newblock \showarticletitle{Deep learning methods in protein structure
  prediction}.
\newblock \bibinfo{journal}{\emph{Comput and Struct Biotech J}}
  \bibinfo{number}{S2001037019304441} (\bibinfo{year}{2020}),
  \bibinfo{pages}{1--10}.
\newblock


\bibitem[\protect\citeauthoryear{Vendruscolo, Kussell, and Domany}{Vendruscolo
  et~al\mbox{.}}{1997}]%
        {VendruscoloDomany97}
\bibfield{author}{\bibinfo{person}{M. Vendruscolo}, \bibinfo{person}{E.
  Kussell}, {and} \bibinfo{person}{E. Domany}.}
  \bibinfo{year}{1997}\natexlab{}.
\newblock \showarticletitle{Recovery of protein structure from contact maps}.
\newblock \bibinfo{journal}{\emph{Folding and Design}} \bibinfo{volume}{1},
  \bibinfo{number}{5} (\bibinfo{year}{1997}), \bibinfo{pages}{295--30}.
\newblock


\bibitem[\protect\citeauthoryear{Walsh, Ba{\`u}, Martin, Mooney, Vullo, and
  Pollastri}{Walsh et~al\mbox{.}}{2009}]%
        {walsh2009ab}
\bibfield{author}{\bibinfo{person}{Ian Walsh}, \bibinfo{person}{Davide
  Ba{\`u}}, \bibinfo{person}{Alberto~JM Martin}, \bibinfo{person}{Catherine
  Mooney}, \bibinfo{person}{Alessandro Vullo}, {and} \bibinfo{person}{Gianluca
  Pollastri}.} \bibinfo{year}{2009}\natexlab{}.
\newblock \showarticletitle{Ab initio and template-based prediction of
  multi-class distance maps by two-dimensional recursive neural networks}.
\newblock \bibinfo{journal}{\emph{BMC structural biology}} \bibinfo{volume}{9},
  \bibinfo{number}{1} (\bibinfo{year}{2009}), \bibinfo{pages}{5}.
\newblock


\bibitem[\protect\citeauthoryear{Wang, Sun, Li, Zhang, and Xu}{Wang
  et~al\mbox{.}}{2017}]%
        {WangXu17}
\bibfield{author}{\bibinfo{person}{S. Wang}, \bibinfo{person}{S. Sun},
  \bibinfo{person}{Z. Li}, \bibinfo{person}{R. Zhang}, {and}
  \bibinfo{person}{J. Xu}.} \bibinfo{year}{2017}\natexlab{}.
\newblock \showarticletitle{Accurate de novo prediction of protein contact map
  by ultra-deep learning model}.
\newblock \bibinfo{journal}{\emph{{PLoS} Comput Biol}}  \bibinfo{volume}{13}
  (\bibinfo{year}{2017}), \bibinfo{pages}{e1005324}.
\newblock


\bibitem[\protect\citeauthoryear{Wu and Zhang}{Wu and Zhang}{2008}]%
        {wu2008comprehensive}
\bibfield{author}{\bibinfo{person}{Sitao Wu} {and} \bibinfo{person}{Yang
  Zhang}.} \bibinfo{year}{2008}\natexlab{}.
\newblock \showarticletitle{A comprehensive assessment of sequence-based and
  template-based methods for protein contact prediction}.
\newblock \bibinfo{journal}{\emph{Bioinformatics}} \bibinfo{volume}{24},
  \bibinfo{number}{7} (\bibinfo{year}{2008}), \bibinfo{pages}{924--931}.
\newblock


\bibitem[\protect\citeauthoryear{Yi, Wu, Gan, Torralba, Kohli, and
  Tenenbaum}{Yi et~al\mbox{.}}{2018}]%
        {yi2018neural}
\bibfield{author}{\bibinfo{person}{Kexin Yi}, \bibinfo{person}{Jiajun Wu},
  \bibinfo{person}{Chuang Gan}, \bibinfo{person}{Antonio Torralba},
  \bibinfo{person}{Pushmeet Kohli}, {and} \bibinfo{person}{Josh Tenenbaum}.}
  \bibinfo{year}{2018}\natexlab{}.
\newblock \showarticletitle{Neural-symbolic vqa: Disentangling reasoning from
  vision and language understanding}. In \bibinfo{booktitle}{\emph{Advances in
  Neural Information Processing Systems}}. \bibinfo{pages}{1031--1042}.
\newblock


\bibitem[\protect\citeauthoryear{You, Ying, Ren, Hamilton, and Leskovec}{You
  et~al\mbox{.}}{2018}]%
        {you2018graphrnn}
\bibfield{author}{\bibinfo{person}{Jiaxuan You}, \bibinfo{person}{Rex Ying},
  \bibinfo{person}{Xiang Ren}, \bibinfo{person}{William~L Hamilton}, {and}
  \bibinfo{person}{Jure Leskovec}.} \bibinfo{year}{2018}\natexlab{}.
\newblock \showarticletitle{Graphrnn: Generating realistic graphs with deep
  auto-regressive models}.
\newblock \bibinfo{journal}{\emph{arXiv preprint arXiv:1802.08773}}
  (\bibinfo{year}{2018}).
\newblock


\bibitem[\protect\citeauthoryear{Zaman, Parthasarathy, and Shehu}{Zaman
  et~al\mbox{.}}{2019}]%
        {ZamanShehuBCB19}
\bibfield{author}{\bibinfo{person}{A. Zaman}, \bibinfo{person}{P.
  Parthasarathy}, {and} \bibinfo{person}{A. Shehu}.}
  \bibinfo{year}{2019}\natexlab{}.
\newblock \showarticletitle{Using Sequence-Predicted Contacts to Guide
  Template-free Protein Structure Prediction}. In \bibinfo{booktitle}{\emph{ACM
  Conf on Bioinf and Comp Biol (BCB)}}. \bibinfo{address}{Niagara Falls, NY},
  \bibinfo{pages}{154--160}.
\newblock


\bibitem[\protect\citeauthoryear{Zaman and Shehu}{Zaman and Shehu}{2019}]%
        {ZamanShehuBMC19}
\bibfield{author}{\bibinfo{person}{A. Zaman} {and} \bibinfo{person}{A. Shehu}.}
  \bibinfo{year}{2019}\natexlab{}.
\newblock \showarticletitle{Balancing multiple objectives in conformation
  sampling to control decoy diversity in template-free protein structure
  prediction}.
\newblock \bibinfo{journal}{\emph{BMC Bioinformatics}} \bibinfo{volume}{20},
  \bibinfo{number}{1} (\bibinfo{year}{2019}), \bibinfo{pages}{211}.
\newblock
\showISSN{1471-2105}
\urldef\tempurl%
\url{https://doi.org/10.1186/s12859-019-2794-5}
\showDOI{\tempurl}


\bibitem[\protect\citeauthoryear{Zhang, Ma, Wang, and Zhou}{Zhang
  et~al\mbox{.}}{2018}]%
        {ZhangZhou18}
\bibfield{author}{\bibinfo{person}{G. Zhang}, \bibinfo{person}{L. Ma},
  \bibinfo{person}{X. Wang}, {and} \bibinfo{person}{X. Zhou}.}
  \bibinfo{year}{2018}\natexlab{}.
\newblock \showarticletitle{Secondary Structure and Contact Guided Differential
  Evolution for Protein Structure Prediction}.
\newblock \bibinfo{journal}{\emph{IEEE/ACM Trans Comput Biol and Bioinf}}
  (\bibinfo{year}{2018}).
\newblock
\urldef\tempurl%
\url{https://doi.org/10.1109/TCBB.2018.2873691}
\showDOI{\tempurl}
\newblock
\shownote{preprint.}


\bibitem[\protect\citeauthoryear{Zhao, Song, and Ermon}{Zhao
  et~al\mbox{.}}{2017}]%
        {zhao2017infovae}
\bibfield{author}{\bibinfo{person}{Shengjia Zhao}, \bibinfo{person}{Jiaming
  Song}, {and} \bibinfo{person}{Stefano Ermon}.}
  \bibinfo{year}{2017}\natexlab{}.
\newblock \showarticletitle{Infovae: Information maximizing variational
  autoencoders}.
\newblock \bibinfo{journal}{\emph{arXiv preprint arXiv:1706.02262}}
  (\bibinfo{year}{2017}).
\newblock


\end{thebibliography}

\section*{Acknowledgements}

This work was supported in part by the National Science Foundation Grant No. 1907805 to AS and LZ and a Jeffress Memorial Trust Award to LZ and AS. The authors additionally thank members of the Zhao and Shehu laboratories for valuable feedback during this work.

\section*{Author contributions}
XG implemented and evaluated the proposed methodologies, as well as drafted the manuscript. ST assisted with preparation of the input data and the evaluation of reconstructed structures. LZ and AS conceptualized the methodologies and provided guidance on implementation and evaluation, as well as edited and finalized the manuscript.

\section*{Competing interests}
The authors declare no competing interests.

\end{document}